\documentclass[a4paper,11pt]{article}
\usepackage{jheppub}
\usepackage[T1]{fontenc}
\usepackage{tensor}
\usepackage{bbm}
\usepackage{parskip}
\usepackage{relsize}
\usepackage{physics}
\usepackage{graphicx}
\usepackage{bigints}
\newcommand{\bb}[1]{\mathbb{#1}}

\newcommand{\hypgeo}[2]{%
  {\vphantom{F}}_{#1}\kern-\scriptspace F_{#2}%
}

\renewcommand{\cal}[1]{\mathcal{#1}}
\renewcommand{\bf}[1]{\mathbf{#1}}

\newcommand{\inv}[1]{\frac{1}{#1}}
\newcommand{\brac}[1]{\left(#1\right)}
\newcommand{\mn}{\mu\nu}

\renewcommand{\dag}{\dagger}
\newcommand{\polylog}[2]{\mathrm{Li}_{#1}\brac{#2}}

\newcommand{\bbm}[1]{\mathbbm{#1}}

\newcommand{\bpsi}{\bar{\psi}}

\newcommand{\cY}{\cal{Y}}
\allowdisplaybreaks

\usepackage{caption}
\usepackage{subcaption}

\usepackage{hyperref}
\title{A Note on the Perturbative Expansion of the Schwinger Model on $S^2$}

\author{Joseph Smith}
\affiliation{School of Mathematics, University of Birmingham \\
Watson Building, Edgbaston, Birmingham B15 2TT, UK}
\emailAdd{j.smith.24@bham.ac.uk}

\abstract{The Schwinger model is perhaps the simplest non-trivial exactly-solvable QFT. In this note we examine the perturbative structure of the theory on the sphere and provide evidence that its quantum corrections match those predicted by the expansion of the exact solution.}

\begin{document} 
\maketitle
\flushbottom

\section{Introduction}
\label{sect: intro}

The Schwinger model, the two-dimensional theory of an Abelian gauge field and massless charged Dirac fermion, has been a source of physical insight since it was first introduced in \cite{Schwinger:1962tp}. The model shares many interesting features with more complex gauge theories, such as colour confinement and quantum mass generation, while remaining simple enough to admit an exact solution. These properties have led to a continued interest in the model and its many generalisations, see for example \cite{Lowenstein:1971fc, Coleman:1975pw, Bergknoff:1976xr, Coleman:1976uz, Roskies:1980jh, Marinari:1981qf, Gass:1982iu, Oki:1984tpr, Jackiw:1984zi, Manton:1985jm, Eller:1986nt, Barcelos-Neto:1986oku, Sachs:1991en, Adam:1993fc, Adam:1996qm, Hosotani:1998za, Fraser-Taliente:2024lea}. Somewhat remarkably, the theory remains exactly solvable when placed on a sphere \cite{Jayewardena:1988td}. In \cite{Anninos:2024fty} it was pointed out that the Lorentzian continuation of this solution gives an exact solution of the Schwinger model on two-dimensional de Sitter spacetime in the Hartle-Hawking vacuum, giving a window into the non-perturbative dynamics of de Sitter QFTs. The study of QFT on de Sitter spacetimes\footnote{See \cite{Bros:1995js, Bros:2010wa, Epstein:2012zz, Anninos:2014lwa, Epstein:2018wfh, Sleight:2019hfp, Isono:2020qew, Pethybridge:2021rwf, Sengor:2021zlc, Penedones:2023uqc, Letsios:2023qzq, Loparco:2023rug, Anninos:2023lin, Loparco:2025azm, Letsios:2025pqo} for an inexhaustive list of references on this topic.} is of considerable interest both from a theoretical and phenomenological perspective; one may hope that solvable models can act as guides to the sort of physics one expects to find in more realistic theories, as well as aiding in the growing effort to understand loop effects in de Sitter QFT  \cite{Anninos:2021ene, Muhlmann:2022duj, Bandaru:2024qvv, Gorbenko:2019rza, LopezNacir:2016gzi, Tsamis:1996qq, Marolf:2010zp, Marolf:2010nz, Anninos:2020hfj, Cacciatori:2024zrv}.

The focus of this note will be the structure of perturbation theory for the Schwinger model on $S^2$. As the exact solution is already known, it may seem somewhat asinine to study the theory perturbatively. However, if we are interested in computing quantum corrections in theories where an exact solution is not available then this viewpoint is flipped: as we have a prediction for the computation of any observable in the theory we can use the Schwinger model as a testing ground for perturbative methods on $S^2$. Our goal in this work is to apply this philosophy by computing the first quantum corrections to the simplest observables in the theory, the partition function and photon propagator\footnote{A comparison of the one-loop fermion propagator diagram and the exact result's expansion was performed in \cite{Anninos:2024fty}, so we shall not focus on this quantity.}. It is also an interesting question to ask to what extent the exact solution can be captured using perturbative methods, as was studied for the flat-space Schwinger model in \cite{Adam:1996qm}. While we won't explore this in the present work, it seems especially pressing for applications to de Sitter QFT where the late-time behaviour of correlation functions differs greatly from perturbative expectations \cite{Anninos:2024fty}.

In the spirit of experimentation we shall calculate both the partition function and photon propagator using two methods. The first employs the position-space representations of propagators and the interaction vertex in stereographic coordinates on the sphere. This has the benefit of having a similar structure to position-space perturbation theory in flat space; however, the resulting integrals we shall find are complicated, and we must be content with a numerical evaluation. The second method is an expansion in angular momentum eigenstates. The difficulty here lies in the structure of the interaction vertex, which is determined by an integral over three of the relevant eigenfunctions. Once this has been ascertained, though, the sums that must be computed for each diagram are relatively simple and can (to an extent) be done explicitly.

The remainder of this work is organised as follows. In section \ref{sect: schwinger review} we review the Schwinger model on $S^2$ and its Feynman rules. In section \ref{sect: stereo pert} we compute the two-loop correction to the partition function and the one-loop correction to the photon propagator using position-space methods, and we follow this in section \ref{sect: ang mom exp} by computing the same objects using the theory's angular momentum expansion. Finally, in section \ref{sect: conclusion} we conclude and discuss potential avenues for future work. We also include two appendices; in appendix \ref{sect: conventions} we set out the conventions we use in this note, and in appendix \ref{sect: gauge anomaly} we discuss the gauge anomaly and how it influences the perturbative calculation of the partition function.

\section{The Schwinger Model on \texorpdfstring{$S^2$}{S2}: A Brief Review} \label{sect: schwinger review}

The Schwinger model consists of a two-dimensional $U(1)$ gauge field coupled to a massless Dirac fermion, with the action
\begin{align} \label{eq: initial action}
    S[A,\bpsi,\psi] &= \bigintsss d^2 x \sqrt{g} \brac{\inv{4} F_{\mn} F^{\mn} + \bpsi \gamma^{\mu} \brac{\nabla_{\mu} + i q A_{\mu} } \psi} \ .
\end{align}
Our fermion conventions are given in appendix \ref{sect: conventions}. We will consider the model on $S^2$, where we will use either standard spherical coordinates or the stereographic coordinate system
\begin{subequations} \label{eq: stereo coords}
\begin{align}
    ds^2 &= \Omega^2 d\bf{x}\cdot d\bf{x} \ , \\
    \Omega &= \frac{2 R^2}{R^2 + \bf{x}\cdot\bf{x}} \ .
\end{align}
\end{subequations}
In order to work with \eqref{eq: initial action} we must choose a gauge-fixing procedure. The most convenient choice is Lorenz gauge,
\begin{equation} \label{eq: gauge fixing}
    d * A = 0 \ ,
\end{equation}
for which we locally have the solution
\begin{equation}
    A = R * d\beta \ ,
\end{equation}
in terms of a prepotential $\beta$, where the factor of $R$ is inserted to keep $\beta$ dimensionless. As we are working with an Abelian gauge theory in Lorenz gauge we do not need to worry about Faddeev-Popov ghosts when gauge-fixing, since they will only contribute to the partition function's overall normalisation. We will assume \eqref{eq: gauge fixing} is obeyed from here onwards. The gauge field's kinetic term is
\begin{align} \nonumber
    F \wedge * F &= R^2 \brac{d * d\beta} \wedge \brac{* d * d\beta} \\ \label{eq: F to phi}
    &= \varepsilon_{S^2} R^2 \brac{\nabla^2 \beta}^2
\end{align}
when written in terms of $\beta$, with the action then
\begin{equation} \label{eq: action in final form}
    S[\beta,\bpsi,\psi] = \bigintsss d^2x \sqrt{g} \brac{
    \frac{R^2}{2} \brac{\nabla^2 \beta}^2 + \bpsi \gamma^{\mu} \nabla_{\mu} \psi + i qR \, \epsilon_{\mn} g^{\nu\rho} \partial_{\rho} \beta\, \bpsi \gamma^{\mu} \psi } \ .
\end{equation}

As we shall be interested in performing perturbative calculations using the action \eqref{eq: action in final form}, it will be useful to review its Feynman rules. We shall first consider perturbation theory in stereographic coordinates, and start by finding the free-field propagators. As this is well-trodden ground we shall be brief in our review. The two-point function of $\beta$,
\begin{equation}
    G(x,y) = \langle \beta(x) \beta(y) \rangle_0 \ ,
\end{equation}
solves the biharmonic Green's function equation
\begin{equation} \label{eq: biharmonic}
    \nabla^4_x G(x,y) = \inv{R^2} \brac{\frac{\delta^{(2)}(x-y)}{\sqrt{g_x}} - \inv{4\pi R^2} } \ ,
\end{equation}
where the constant offset to the delta function cancels the zero-mode piece in its spherical harmonic expansion. This is most easily solved by expanding both sides in spherical harmonics (for which our conventions are given in appendix \ref{sect: conventions}), where we see that
\begin{equation} \label{eq: G ang mom expansion}
    G(x,y) = \inv{R^2} \sum_{l=1}^{\infty} \sum_{m=-l}^l \frac{\cal{Y}_{lm}(x) \cal{Y}_{lm}(y)}{l^2(l+1)^2} + c_0 \ ,
\end{equation}
in terms of an arbitrary constant $c_0$. This constant is determined by the zero-mode of $\beta$, which we shall take to vanish. This is then nothing more than the expansion of
\begin{equation} \label{eq: free phi 2-point}
    G(x,y) = \inv{4\pi} \brac{
    \polylog{2}{1 - \frac{u(x,y)}{2}}
    + 1 - \frac{\pi^2}{6} } \ ,
\end{equation}
where $u(x,y)$ is the $SO(3)$-invariant function defined in stereographic coordinates by
\begin{equation}
    u(\bf{x},\bf{y}) = \frac{2 R^2 (\bf{x}-\bf{y})\cdot (\bf{x}-\bf{y})}{(R^2 + \bf{x} \cdot \bf{x})(R^2 + \bf{y}\cdot \bf{y} )} \ .
\end{equation}
This function is finite as we take the two points to coincide, with the value
\begin{equation} \label{eq: G at coincident points}
    G(x,x) = \inv{4\pi} \ .
\end{equation}

Next, let us consider the fermion's propagator. This is simplified by making the field reparameterisation
\begin{equation} \label{eq: flat fermion}
    \psi = \Omega^{-1/2} \chi \ ,
\end{equation}
which transforms the curved-space fermion's action into that of a free massless fermion on the flat two-dimensional plane,
\begin{equation}
    \int d^2x \sqrt{g} \, \bpsi \gamma^{\mu} \nabla_{\mu} \psi = \int d^2 x \, \bar{\chi} \pmb{\sigma}^a \partial_a \chi \ .
\end{equation}
The two-point function of $\psi$ in stereographic coordinates is then
\begin{align} \nonumber
    S(\bf{x},\bf{y}) &= \langle \psi(\bf{x}) \bpsi(\bf{y}) \rangle_0 \\ \nonumber
    &= \inv{\sqrt{\Omega(\bf{x}) \Omega(\bf{y})}} \langle \chi(\bf{x}) \bar{\chi}(\bf{y}) \rangle_0 \\
    &= \inv{2\pi\sqrt{\Omega(\bf{x}) \Omega(\bf{y})}} \frac{\pmb{\sigma}^a (\bf{x}-\bf{y})^a}{(\bf{x} - \bf{y}) \cdot (\bf{x} - \bf{y}) } \ ,
\end{align}
where we've used the well-known propagator for a flat free-fermion in two-dimensions. When needed we will use the notation
\begin{equation}
    S(x,y) = \frac{S_0(x,y)}{\sqrt{\Omega(x)\Omega(y)}} 
\end{equation}
to isolate the non-flat contribution to the propagator.

Perturbative computations can then be performed using Feynman diagrams, with the propagators
\begin{subequations}
\begin{align}
    \begin{gathered}
        \includegraphics[width=0.17\linewidth]{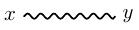}
    \end{gathered} &= G(x,y) \ , \\[-0.2em]
    \begin{gathered}
        \includegraphics[width=0.17\linewidth]{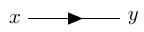}
    \end{gathered} &= S(x,y) \ ,
\end{align}
\end{subequations}
for the prepotential and fermion. At each interaction vertex we connect these using
\begin{equation}
    \begin{gathered}
        \includegraphics[width=0.185\linewidth]{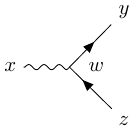}
    \end{gathered} = -i q R \epsilon_{\mn} g^{\nu\rho} \partial_{\rho}^{(w)} G(x,w) S(z,w) \gamma^{\mu}(w) S(w,y) \ ,
\end{equation}
before integrating over all internal points. As usual, we must also account for minus signs from fermion loops and symmetry factors. If we work in stereographic coordinates, this simplifies nicely; $\epsilon_{\mn}, g_{\mn}$, and $\gamma^{\mu}$ all become their flat space counterparts, $S(x,y)$ is replaced with $S_0(\bf{x},\bf{y})$, and the integrals over $S^2$ become integrals over $\bb{R}^2$. In this parametrisation of the Feynman integrals, the only place we see the spherical geometry is in the two-point function of $\beta$.

An alternative to the method described above is to utilise the structure of the sphere and perform an expansion in eigenfunctions of the Laplacian and Dirac operator. For the prepotential, this means an expansion in real spherical harmonics,
\begin{equation}
    \beta(x) = \sum_{l,m}  b_{lm} \cal{Y}_{lm}(x) \ .
\end{equation}
Using \eqref{eq: G ang mom expansion} it is straightforward to see that the free-field two-point function is 
\begin{equation}
    \big\langle b_{l_1m_1} b_{l_2m_2} \big\rangle_0 = \frac{\delta_{l_1 l_2} \delta_{m_1 m_2}}{l_1^2 (l_1+1)^2} \equiv P_{l_1} \delta_{l_1 l_2} \delta_{m_1 m_2} \ .
\end{equation}
Similarly, we can expand $\psi$ in eigenfunctions of the Dirac operator\footnote{The details of these functions are given in appendix \ref{sect: conventions}.} \cite{Camporesi:1995fb},
\begin{equation}
    \psi(x) = \sqrt{R} \sum_{L=0}^{\infty} \sum_{M=0}^L \sum_{\sigma,s=\pm1} c_{LM}^{s \sigma} \psi_{LM}^{s \sigma}(x) \ ,
\end{equation}
where $c^{s\sigma}_{LM}$ are Grassmann-valued variables. As the eigenfunctions obey
\begin{equation}
    \gamma^{\mu} \nabla_{\mu} \psi^{s\sigma}_{LM} = \frac{i \sigma (L+1)}{R} \psi^{s \sigma}_{LM} \ ,
\end{equation}
the momentum-space variables have the two-point function
\begin{equation}
     \big\langle c^{s_1 \sigma_1}_{L_1 M_1}  \bar{c}^{s_2 \sigma_2}_{L_2 M_2} \big\rangle_0 = -\frac{\delta_{L_1 L_2} \delta_{M_1 M_2} \delta_{s_1 s_2} \delta_{\sigma_1 \sigma_2} }{i \sigma_1 \brac{L_1 + 1}} \equiv S_{L_1 \sigma_1} \delta_{L_1 L_2} \delta_{M_1 M_2} \delta_{s_1 s_2} \delta_{\sigma_1 \sigma_2}  \ .
\end{equation}

Finally, we need to deal with the interaction term in \eqref{eq: action in final form}. Integrating it by parts gives
\begin{equation}
    S_{int} = q R \int d^2x \, \sqrt{g} \, \beta \brac{
    \bpsi \gamma_* \gamma^{\mu} \nabla_{\mu} \psi - \nabla_{\mu} \bpsi \gamma^{\mu} \gamma_* \psi } \ ,
\end{equation}
which can be expanded in harmonics to get
\begin{align} \nonumber
    S_{int} &= i q R \sum_{lm} \sum_{L_i, M_i, s_i, \sigma_i} b_{lm} \bar{c}^{s_1 \sigma_1}_{L_1 M_1} c_{L_2 M_2}^{s_2 \sigma_2}  \bigg[ \sigma_1 \brac{L_1 +1} + \sigma_2 \brac{L_2 + 1} \bigg] \int d^2x \sqrt{g} \, \cY_{lm} \psi^{s_1 \sigma_1 \, \dag}_{L_1 M_1} \gamma_* \psi^{s_2 \sigma_2}_{L_2 M_2} \\
    &\equiv i q R \sum_{lm} \sum_{L_i, M_i, s_i, \sigma_i} b_{lm} \bar{c}^{s_1 \sigma_1}_{L_1 M_1} c_{L_2 M_2}^{s_2 \sigma_2}  \bigg[ \sigma_1 \brac{L_1 +1} + \sigma_2 \brac{L_2 + 1} \bigg]  \cal{A}^{s_1 \sigma_1 s_2 \sigma_2}_{l m L_1 M_1 L_2 M_2} \ .
\end{align}
The important part of this are the integrals
\begin{equation}
    \cal{A}^{s_1 \sigma_1 s_2 \sigma_2}_{l m L_1 M_1 L_2 M_2} = \int d^2x \sqrt{g} \, \cY_{lm} \psi^{s_1 \sigma_1 \, \dag}_{L_1 M_1} \gamma_* \psi^{s_2 \sigma_2}_{L_2 M_2} \ ,    
\end{equation}
over the different harmonics. A complete characterisation of the momentum-space interaction vertex requires an expression for these; we shall return to this point in section \ref{sect: ang mom exp}.

We can now state the momentum-space Feynman rules. For convenience we shall use $\rho_i= (L_i,M_i,s_i,\sigma_i)$ as a multi-index for our fermion labels. We assign the propagators
\begin{subequations}
\begin{align}
    \begin{gathered}
        \includegraphics[width=0.28\linewidth]{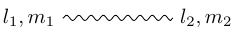}
    \end{gathered} &= \delta_{l_1 l_2} \delta_{m_1 m_2} P_{l_1} \ , \\
    \begin{gathered}
        \includegraphics[width=0.23\linewidth]{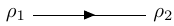}
    \end{gathered} &= \delta_{\rho_1 \rho_2} S_{L_1 \sigma_1} \ , 
\end{align}
\end{subequations}
to each line in a diagram, and for each vertex we include the factor
\begin{equation} \label{eq: interaction vertex}
\begin{gathered}
    \includegraphics[width=0.22\linewidth]{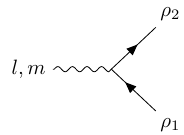}
\end{gathered} = - i q R \Big( \sigma_1 \brac{L_1 +1} + \sigma_2 \brac{L_2 + 1} \Big)  \cal{A}_{lm \rho_1 \rho_2}  \ .
\end{equation}
Note that the eigenspinor index $\sigma_i$ used in the sphere expansion should not be confused with the Pauli matrices $\pmb{\sigma}^a$, which will always be written in bold to properly differentiate the two. We then sum over all momentum labels and include symmetry factors and fermion-loop minus sign contributions.

\section{Perturbation Theory in Stereographic Coordinates} \label{sect: stereo pert}

\subsection{Two-Loop Partition Function and Regularisation}
\label{sect: partition function}

The first object we will be interested in computing is the first quantum correction to the sphere partition function, defined by
\begin{equation} \label{eq: partition function}
    \cal{Z}_q = \bigintsss \frac{DA D\bpsi D\psi}{\text{vol}(\cal{G})}  \, e^{-S[A,\bpsi,\psi]} \ ,
\end{equation}
where $\text{vol}(\cal{G})$ is the volume of the gauge group, or, after gauge-fixing using \eqref{eq: gauge fixing},
\begin{align} \label{eq: partition function with Phi}
    \cal{Z}_q = \bigintsss D\beta D\bpsi D\psi \, e^{-S[\beta,\bpsi,\psi]} \ .
\end{align}
This was computed exactly in \cite{Anninos:2024fty}; one can then expand it perturbatively in $qR$, where we find the first correction
\begin{align} \label{eq: partition function exact result}
    \ln\brac{\frac{\cal{Z}_q}{\cal{Z}_0}}\bigg\rvert_{q^2 R^2} = \frac{q^2 R^2 \brac{2\ln\epsilon + 1 - 2\gamma}}{2\pi} + O(\epsilon) \ .
\end{align}
Here $\epsilon \sim \brac{\Lambda_{UV} R}^{-1}$ is a UV cut-off. The finite terms in this expression can be altered by a rescaling of the cut-off, so have no physical meaning; we will therefore focus on recovering the coefficient of the logarithm. 

\begin{figure}
    \centering
    \includegraphics[width=0.8\linewidth]{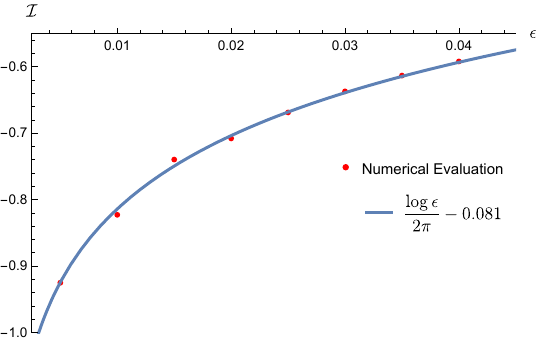}
    \caption{Numerical evaluation of the leading quantum correction to the partition function \eqref{eq: integral} using the minimum length regularisation \eqref{eq: min length reg}.}
    \label{fig: Min Length Regularisation Partition Function}
\end{figure}

We would like to recover this by directly performing an asymptotic expansion of \eqref{eq: partition function with Phi}, leading us to the perturbative expression
\begin{align} \nonumber
    \cal{Z}_q \sim \cal{Z}_0 \,\mathlarger{\sum}_{n=0}^{\infty} \frac{(-iqR)^n}{n!} 
    &\bigintsss d^2 x_1 ... d^2 x_n \, \sqrt{g_1 ... g_n}\,  \epsilon_{\mu_1 \nu_1} g^{\nu_1 \rho_1}(x_1) \,...\, \epsilon_{\mu_n \nu_n} g^{\nu_n \rho_n}(x_n)
    \\
    &\times \left\langle 
    \brac{\partial_{\rho_1} \beta \,\bpsi \gamma^{\mu_1} \psi}(x_1) \, ...\, \brac{\partial_{\rho_n} \beta \,\bpsi \gamma^{\mu_n} \psi}(x_n)
    \right\rangle_0 \ ,
\end{align}
where $\cal{Z}_0$ is the partition function of the free theory. This becomes simpler if we work in the stereographic coordinate system \eqref{eq: stereo coords} and use the conformally rescaled fermion \eqref{eq: flat fermion}, where we find the partition function
\begin{align}
    \cal{Z}_q \sim \cal{Z}_0 \,\mathlarger{\sum}_{n=0}^{\infty} \frac{(-iqR)^n}{n!} 
    &\bigintsss d^2 \bf{x}_1 ... d^2 \bf{x}_n \,  \epsilon_{a_1 b_1}  \,...\, \epsilon_{a_n b_n}
    \\
    &\times \left\langle 
    \brac{\partial_{b_1} \beta \,\bar{\chi} \pmb{\sigma}^{a_1} \chi}(\bf{x}_1) \, \,...\, \brac{\partial_{b_n} \beta \,\bar{\chi}\pmb{\sigma}^{a_n}  \chi}(\bf{x}_n)
    \right\rangle_0 \ .
\end{align}
When computing the correlation function's Wick contractions, we will assume that we are always working with a regularisation scheme for which $\tr\brac{\sigma^a S(x,x)} =0$, so any diagrams containing
\begin{equation}
\begin{gathered}
    \includegraphics[width=0.25\textwidth]{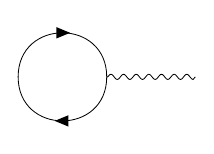}
\end{gathered}
\end{equation} 
as a subdiagram vanish. Taking the logarithm of both sides, we see that the leading-order correction to the sphere partition function is\footnote{From here onwards we will neglect the distinction between an asymptotic and convergent expansion.}
\begin{equation}
    \ln\brac{\frac{\cal{Z}_q}{\cal{Z}_0}}\bigg\rvert_{q^2 R^2} = \frac{q^2 R^2}{2} \bigintsss d^2 \bf{x} d^2 \bf{y} \, \epsilon_{ac} \epsilon_{bd} \brac{\partial_c^{(\bf{x})} \partial_d^{(\bf{y})} G(\bf{x},\bf{y}) } \tr\brac{
    \pmb{\sigma}^a S_0(\bf{x},\bf{y}) \pmb{\sigma}^b S_0(\bf{y},\bf{x})} \ ,
\end{equation}
which is the contribution associated with the vacuum bubble diagram
\begin{equation} \label{eq: two-loop partition function diagram}
\begin{gathered}
    \includegraphics[width=0.16\textwidth]{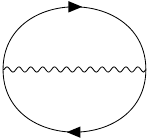}
\end{gathered} \ .
\end{equation}
As $G(\bf{x},\bf{y})$ only appears differentiated here, we are free to shift it by a constant while leaving the integral unchanged. We can then replace it by
\begin{equation}
    G(\bf{x},\bf{y}) \to \Tilde{G}(\bf{x},\bf{y}) = G(\bf{x},\bf{y}) - \inv{4\pi} \ ,
\end{equation}
which using \eqref{eq: G at coincident points} the property that
\begin{equation}
     \Tilde{G}(\bf{x},\bf{x}) = 0 \ .
\end{equation}
Integrating by parts, the integral we are interested in evaluating is therefore
\begin{equation} \label{eq: integral}
    \cal{I} = \inv{2} \bigintsss d^2\bf{x} d^2\bf{y}  \, \Tilde{G}(\bf{x},\bf{y}) \, \epsilon_{ac} \epsilon_{bd} \partial_c^{(\bf{x})} \partial_d^{(\bf{y})} \tr \brac{\pmb{\sigma}^a S_0(\bf{x},\bf{y}) \pmb{\sigma}^b S_0(\bf{y},\bf{x}) } \ .
\end{equation}
As expected, this integral is ill-defined due to divergences at coincident points and must be regulated. Since $\Tilde{G}(\bf{x},\bf{y})$ is finite as $\bf{x}\to \bf{y}$, the divergence comes from the fermion loop contribution
\begin{equation} \label{eq: definition of L}
    L^{ab}(\bf{x},\bf{y}) = \tr \brac{\pmb{\sigma}^a S_0(\bf{x},\bf{y}) \pmb{\sigma}^b S_0(\bf{y},\bf{x}) } \ .
\end{equation}
However, we must be careful about our choice of regularisation. As discussed in appendix \ref{sect: gauge anomaly}, if we do not choose a gauge-invariant regularisation of $L^{ab}$ then the coefficient of the logarithm will be reduced and we will not recover \eqref{eq: partition function exact result}. For instance, suppose we chose to regulate \eqref{eq: definition of L} by introducing a cut-off lengthscale $\epsilon R$ in $S_0$, i.e.
\begin{equation} \label{eq: min length reg}
    S_0(\bf{x},\bf{y}) \to \Tilde{S}_0(\bf{x},\bf{y}) = \frac{\pmb{\sigma}^a (\bf{x}-\bf{y})^a}{2\pi \brac{(\bf{x} - \bf{y}) \cdot (\bf{x} - \bf{y}) + \epsilon^2 R^2 }} \ .
\end{equation}
This generates a gauge anomaly, which we show explicitly in appendix \ref{sect: gauge anomaly}, and so we expect the integral \eqref{eq: integral} to give a logarithmic divergence whose coefficient is half the correct result. This is exactly the case, as seen in figure \ref{fig: Min Length Regularisation Partition Function}.

Instead, we will employ a Pauli-Villars (PV) regularisation scheme. We shall, however, be somewhat unconventional. As we saw above, when working in stereographic coordinates with the rescaled fermions $\eqref{eq: flat fermion}$ the only place that we see the spherical geometry is in the propagator of $\beta$. Since we only need to regularise the fermion loop $L^{ab}$ any gauge-invariant flat-space regularisation scheme will extend to a gauge-invariant regularisation of the full theory. For numerical convenience we will therefore introduce flat-space PV regulator fields; from the perspective of the sphere theory these are fields with spatially-varying masses and the action
\begin{equation}
    S_{PV}[\rho_i] = \mathlarger{\sum}_{i=1}^N \bigintsss d^2 x \sqrt{g} \,
    \bar{\rho}_i \bigg( \gamma^{\mu} \nabla_{\mu} + i q \gamma^{\mu} A_{\mu} + M_i(x,\epsilon)  \bigg) \rho_i \ ,
\end{equation}
where
\begin{equation}
    M_i(x,\epsilon) = \frac{c_i}{\epsilon \,  \Omega^2(x) R} \equiv \frac{m_i(\epsilon)}{\Omega^2(x)} \ ,
\end{equation}
where we may always take $c_i>0$ through a redefinition of the regulator fields. We decouple the additional fields and recover the original theory when $\epsilon$ is taken to vanish. By making the field redefinition
\begin{equation}
    \rho_i = \frac{\lambda_i}{\sqrt{\Omega(x)}}  \ ,
\end{equation}
we see that the action takes the nicer form
\begin{equation}
    S_{PV}[\lambda_i] = \mathlarger{\sum}_{i=1}^N \bigintsss d^2 \bf{x} \, \bar{\lambda}_i \bigg( \pmb{\sigma}^a \partial_a + i q R \,\varepsilon_{ab} \pmb{\sigma}^a \partial_b \beta + m_i(\epsilon)  \bigg) \lambda_i \ ,
\end{equation}
in stereographic coordinates. We shall take the fields to be commuting for odd values of $i$ and anticommuting for even values, and always introduce an odd number of PV fields so that (after including the physical fermion) each anticommuting field is paired with a commuting field.

The free propagators of the regulator fields are
\begin{align} \nonumber
    \langle \lambda_i(\bf{x}) \bar{\lambda}_i(\bf{y}) \rangle_0 &= \frac{m_i}{2\pi} \brac{
    K_0(m_i |\bf{x} - \bf{y}|) \bbm{1}_2 - \frac{\pmb{\sigma}^a (x-y)_a}{|\bf{x} - \bf{y}|} K_{-1}(m_i |\bf{x} - \bf{y}|)
    } \\
    &\equiv S_0^{\epsilon_i}
\end{align}
The regularised fermion loop is then
\begin{equation} \label{eq: regulated loop}
    L_{reg.}^{ab}(\bf{x},\bf{y}) = L^{ab}(\bf{x},\bf{y}) + \sum_{i=1}^N (-1)^i L_{\epsilon_i}^{ab}(\bf{x},\bf{y}) \ ,
\end{equation}
where $L^{ab}_{\epsilon_i}$ is defined by replacing $S_0$ with $S_0^{\epsilon_i}$ in $L^{ab}$. While a single regulator field suffices to regulate the fermion loop, once we include the internal prepotential propagator we must include additional PV fields. In section \ref{sect: ang mom exp} we shall see that one can regulate the integral by also adding a single prepotential regulator field (as is typically done when working in flat spacetimes); however, here this leads to unsatisfactory numerical results for the partition function. A solution that yields more precise results is to solely use fermionic PV fields, where to ensure numerical accuracy we ask for a set of coefficients $c_i$ for which \eqref{eq: regulated loop} has no divergences as we take $\bf{x}\to\bf{y}$\footnote{This is a sufficient condition for a choice of regulator, though by no means necessary.}. The first value of $N$ for which we can do this is $N=5$.

\begin{figure}
    \centering
    \includegraphics[width=0.8\linewidth]{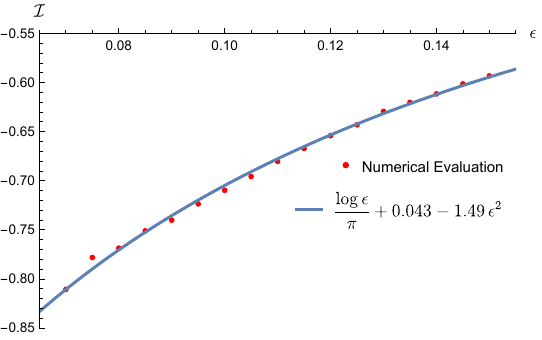}
    \caption{Numerical evaluation of the leading quantum correction to the partition function \eqref{eq: integral} using the Pauli-Villars regularisation \eqref{eq: regulated loop} using the parameters \eqref{eq: c2 and c3 2} and \eqref{eq: c4 and c5 2}.}
    \label{fig: PV Regularisation Partition Function}
\end{figure}

By rescaling $\epsilon$ we are always free to take $c_1 = 1$, which we shall do from here onwards. As everything in $L^{ab}_{reg.}$ is both translationally and rotationally invariant, we can analyse the divergences by considering the function for $\bf{x} = (x,0)$ (with $x>0$) and $\bf{y}=0$. As we take $x\to0$ the off-diagonal terms vanish, and for the diagonal terms we find
\begin{align} \nonumber
    L_{reg.}^{aa}(x) = \,& \inv{2\pi^2 \epsilon^2 R^2} \ln\brac{\frac{x}{2\epsilon R}}^2 \brac{ c_2^{-2} - c_3^{-2} + c_4^{-2} - c_5^{-2} - 1 } \\ \nonumber
    &+ \inv{4\pi^2 \epsilon^2 R^2} \ln\brac{\frac{x}{2\epsilon R}} \bigg[ 2(2\gamma + (-1)^a)\brac{ c_2^{-2} - c_3^{-2} + c_4^{-2} - c_5^{-2} - 1 } \\
    &- 4 \brac{ c_2^{-2} \ln c_2 - c_3^{-2} \ln c_3 + c_4^{-2} \ln c_4 - c_5^{-2} \ln c_5 } \bigg] + O(1) \ .
\end{align}
If we imagine fixing $c_2$ and $c_3$, we can remove all divergences in our theory by taking $c_5$ to be
\begin{equation} \label{eq: c5 eq}
    c_5 = \frac{c_2 c_3 c_4}{\sqrt{c_3^2 c_4^2 - c_2^2 c_4^2 + c_2^2 c_3^2 - c_2^2 c_3^2 c_4^2}} \ ,
\end{equation}
and $c_4$ to obey the transcendental equation
\begin{equation} \label{eq: trans equation}
    c_3^2 c_4^2 c_5^2 \ln c_2 - c_2^2 c_4^2 c_5^2 \ln c_3 + c_2^2 c_3^2 c_5^2 \ln c_4 - c_2^2 c_3^2 c_4^2 \ln c_5 = 0 \ .
\end{equation}
A consistent solution can be found using the values
\begin{equation} \label{eq: c2 and c3 2}
    c_2 = \frac{22414378}{24269829} \quad , \quad c_3 = \frac{127 056 109}{34130186} \ ,
\end{equation}
where $c_4$ and $c_5$ have the approximate forms
\begin{equation} \label{eq: c4 and c5 2}
    c_4 = 0.252949... \quad , \quad c_5 = 0.252141... \ .
\end{equation}
While $L^{ab}_{reg.}$ is non-divergent, we still need to see if the integral \eqref{eq: integral} is finite with this regularisation; in particular, we must check whether differentiating the function is well-defined. It turns out that $L^{ab}_{reg.}$ differentiable everywhere except at coincident points; however, in the integral such points are naturally absent as they are where $\Tilde{G}$ vanishes, and so the integrand is finite at all points. We can therefore compute the variation of the sphere partition function with $\epsilon$, obtaining the result in figure \ref{fig: PV Regularisation Partition Function}. Due to the more complicated nature of the regularisation, we are unable to go to as small values of $\epsilon$ as we could for the minimum length regularisation \eqref{eq: min length reg} due to increasing numerical error. This means that we should generically expect contributions from the higher order terms in $\epsilon$ to play a role in the regulated partition function. Assuming an expansion in even powers as the sign of the regulator mass should not play a role, our numerical result should be of the form
\begin{equation}
    \cal{I}^{(\epsilon)}_{PV} = \alpha \ln \epsilon + \beta + \gamma \epsilon^2 + O(\epsilon^4) \ ,
\end{equation}
where we would expect that $\beta,\gamma$ are at most $O(1)$ coefficients. We get excellent agreement with the function
\begin{equation}
    f(\epsilon) = \frac{1}{\pi} \ln \epsilon + 0.043 - 1.49 \epsilon^2 \ ,
\end{equation}
which is consistent with the exact result. As hoped, both coefficients are $O(1)$.

\begin{figure}
    \centering
    \includegraphics[width=0.8\linewidth]{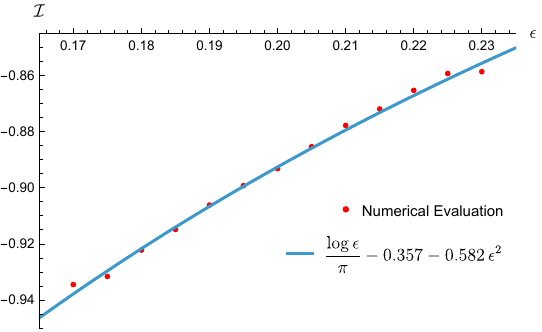}
    \caption{Numerical evaluation of the leading quantum correction to the partition function \eqref{eq: integral} using the Pauli-Villars regularisation \eqref{eq: regulated loop} using the parameters \eqref{eq: c2 and c3} and \eqref{eq: c4 and c5}.}
    \label{fig: PV Regularisation Partition Function 2}
\end{figure}

We should check that the prior result for the logarithmic divergence's coefficient is independent of the precise choice of PV regularisation we use; to do this, we will perform the computation again for another set of parameters $c_i$. Another solution to \eqref{eq: c5 eq} and \eqref{eq: trans equation} is given by the values
\begin{equation} \label{eq: c2 and c3}
    c_2 = \frac{3}{16} \quad , \quad c_3 = \frac{5}{16} \ ,
\end{equation}
where $c_4$ and $c_5$ have the approximate forms
\begin{equation} \label{eq: c4 and c5}
    c_4 = 0.255818... \quad , \quad c_5 = 0.175452... \ .
\end{equation}
The numerical accuracy obtained using these coefficients is slightly worse than in the previous case and we must work at higher values of $\epsilon$, giving the result shown in figure \ref{fig: PV Regularisation Partition Function 2}. We see that the numerical evaluation fits well with the function
\begin{equation}
    f(\epsilon) = \frac{1}{\pi} \ln \epsilon - 0.357-0.582 \epsilon^2   \ .
\end{equation}
This is again consistent with the exact result for the regulator-independent logarithmic divergence, though of course the finite terms differ between the two regularisations. 

To summarise, we have found that (as could have been expected) we only reproduce the exact result if we work with a gauge-invariant regularisation scheme. However, using regulators that induce a gauge-anomaly allows for more numerically efficient computations. This gives another way to compute quantities provided one accounts for the anomalous factor in front of the diagram, which is easily deduced; we shall see an example of this below.

\subsection{One-Loop Contribution to the Photon Propagator}
\label{sect: propagator}

A related observable that we can compute is the two-point function of the prepotential,
\begin{equation} \label{eq: two-point function def}
    \langle\beta(x) \beta(y) \rangle = \inv{\cal{Z}_q} \bigintsss D\beta D\bpsi D\psi \, \beta(x) \beta(y) \, e^{-S[\beta,\bpsi,\psi]} \ .
\end{equation}
In \cite{Anninos:2024fty} this was computed exactly, giving the result\footnote{There is a slight difference in our normalisation of the prepotential compared to the one used there, meaning the two-point function here differs by a factor of $(qR)^{-2}$.}
\begin{align} \nonumber
    \langle \beta(x) \beta(y) \rangle = \, &- \inv{4 q^2 R^2} \bigg(
    1 - \frac{\pi}{q^2 R^2} + \ln\frac{u(x,y)}{2} \\
    &+ \Gamma(\Delta) \Gamma(1-\Delta) \hypgeo{2}{1}\brac{\Delta, 1-\Delta, 1,1- \frac{u(x,y)}{2}} \bigg)
\end{align}
with
\begin{equation}
    \Delta = \frac{1}{2} + i \sqrt{\frac{q^2 R^2}{\pi} - \inv{4}} \ .
\end{equation}
After expanding this in powers of $qR$ we see that the $q$-independent terms are just the free-field two-point function \eqref{eq: free phi 2-point}, and the leading-order corrections to this are given by
\begin{align} \nonumber
    \langle \beta(x) \beta(y) \rangle \bigg{\rvert}_{q^2 R^2} = \frac{q^2 R^2}{4\pi^2} \Bigg[ &
    2 - \frac{\pi^2}{6} + \ln\frac{u(x,y)}{2} \Bigg(
    \ln\frac{u(x,y)}{2} \ln\brac{ 1 - \frac{u(x,y)}{2} }
    \\ \nonumber
    &+ \polylog{2}{1 - \frac{u(x,y)}{2}} - \frac{\pi^2}{6} \Bigg) + 
    \polylog{2}{1 - \frac{u(x,y)}{2}} \\ \label{eq: two-point function at one-loop}
    &+ 2 \polylog{3}{\frac{u(x,y)}{2}} - 2 \zeta(3) \Bigg] \ .
\end{align}

\begin{figure}
    \centering
    \includegraphics[width=0.8\linewidth]{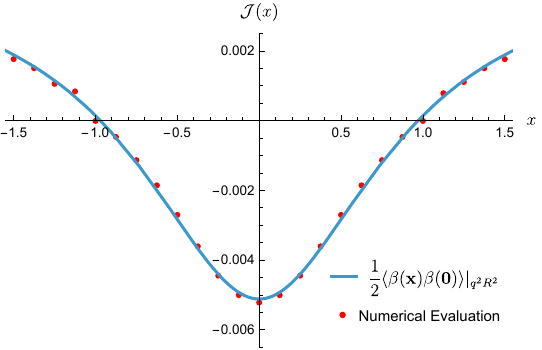}
    \caption{Numerical evaluation of the one-loop correction to the prepotential two-point function \eqref{eq: one-loop prepotential diagram} using the minimum length regularisation \eqref{eq: min length reg} and \eqref{eq: min length u reg} with $\epsilon = 1/2000$.}
    \label{fig: Min Length One Loop}
\end{figure}

Let us now compute the same term in perturbation theory. A direct expansion of \eqref{eq: two-point function def} yields the terms
\begin{align} \nonumber
    \langle \beta(x) \beta(y) \rangle = G(x,y) \, + \, &q^2 R^2 \bigintsss d^2 z d^2 w \, \sqrt{g_z g_w}\, \epsilon_{\mu_1 \nu_1} g^{\nu_1 \rho_1}\epsilon_{\mu_2 \nu_2} g^{\nu_2 \rho_2}\, \partial_{\rho_1}^{(z)} G(x,z) \\
    &\times\partial_{\rho_2}^{(w)} G(w,y)  \, \tr\brac{\gamma^{\mu}(z) S(z,w) \gamma^{\nu}(w) S(w,z)  } + O(q^4 R^4) \ .
\end{align}
We are interested in the integral, which arises from the one-loop diagram
\begin{equation}
\begin{gathered}
    \includegraphics[width = 0.33\linewidth]{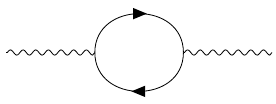}
\end{gathered} \qquad .
\end{equation} 
In stereographic coordinates this is just
\begin{align} \label{eq: one-loop prepotential diagram}
    \mathcal{J}(\bf{x}, \bf{y}) =\, q^2 R^2 \int d^2 \bf{z} d^2 \bf{w} \, \epsilon_{ac} \epsilon_{bd} \, \partial_c^{(\bf{z})} G(\bf{x}, \bf{z}) \partial_d^{(\bf{w})} G(\bf{w}, \bf{y})  L^{ab}(\bf{z},\bf{w})  \ ,
\end{align}
and should agree with \eqref{eq: two-point function at one-loop} if a non-anomalous regularisation is used. In fact, to aid with the numerical evaluation we shall use the minimum-length regularisation \eqref{eq: min length reg} for the fermions and so expect the result to be half \eqref{eq: two-point function at one-loop}. We shall also not integrate by parts, meaning we must also regulate divergences in derivatives of the prepotential propagators in the same manner, taking
\begin{equation} \label{eq: min length u reg}
    u(\bf{x}, \bf{y}) \to u_{\epsilon}(\bf{x}, \bf{y}) = \frac{2 R^2 \Big( (\bf{x}-\bf{y})\cdot (\bf{x}-\bf{y}) + \epsilon^2 R^2 \Big)}{(R^2 + \bf{x} \cdot \bf{x})(R^2 + \bf{y}\cdot \bf{y} )} \ .
\end{equation}
The result of this is plotted in figure \ref{fig: Min Length One Loop} for $\bf{x} = (x,0)$ and $\bf{y} = \bf{0}$, which we see is consistent with the predicted curve.

\section{Perturbation Theory Using an Angular Momentum Expansion} \label{sect: ang mom exp}

\subsection{One-Loop Vacuum Polarisation and the Photon Propagator}

We can now revisit the previous calculations using the angular momentum expansion. The most interesting quantity we shall consider from this perspective is the exact two-point function of the prepotential. The result of quantum effects is the inclusion of a mass term for the prepotential, and so we find
\begin{align} \nonumber
    \big\langle b_{l_1 m_1} b_{l_2 m_2} \big\rangle &= \frac{\delta_{l_1  l_2} \delta_{m_1 m_2}}{l_1^2 (l_1 + 1)^2 + \frac{q^2 R^2}{\pi}l_1 (l_1 + 1)} \\ \label{eq: two point function momentum space exact}
    &\equiv \big\langle b_{l_1 m_1} b_{l_2 m_2} \big\rangle_0 \bigg( 1 + \frac{q^2 R^2}{l_1 (l_1 + 1)\pi } \bigg)^{-1}
\end{align}
in momentum space. In perturbation theory the relevant diagram for computing this at one-loop is the vacuum polarisation 
\begin{equation} \label{eq: vac pol diagram}
\begin{gathered}
    \includegraphics[width = 0.27\textwidth]{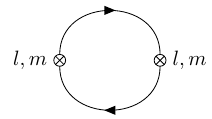}
\end{gathered}
 = - \frac{q^2 R^2}{2} A_{lm}  \ ,
\end{equation}
where the vertices denote that we are including the interaction \eqref{eq: interaction vertex} without an external prepotential propagator. It is well-known from the non-perturbative solution of the model that the prepotential's mass is controlled by the one-loop exact chiral anomaly, meaning repeated sequential insertions of \eqref{eq: vac pol diagram} will recover all terms in the series expansion of \eqref{eq: two point function momentum space exact}. Here $SO(3)$-covariance allows us to impose momentum conservation, forcing $l$ and $m$ to be equal at both vertices. In fact, the symmetry also requires that $A_{lm}$ is independent of $m$, which we shall use in its calculation. Using the Feynman rules outlined above it is straightforward to see that
\begin{equation} \label{eq: Alm def}
    A_{lm} = -\mathlarger{\sum}_{\rho_i} \Big(\sigma_1 (L_1 + 1) + \sigma_2 (L_2 + 1) \Big)^2 S_{L_1\sigma_1} S_{L_2 \sigma_2} \cal{A}_{lm \rho_1 \rho_2} \cal{A}_{lm \rho_2 \rho_1} \ .
\end{equation}
The only diagrams that contribute to the prepotential's two-point function are increasing numbers of insertions of the diagram \eqref{eq: vac pol diagram}\footnote{This is just the statement that quantum corrections are governed by the two-dimensional chiral anomaly, which is one-loop exact.} \cite{Adam:1996qm}, and performing the resulting geometric sum gives
\begin{equation}
    \big\langle b_{l_1 m_1} b_{l_2 m_2} \big\rangle = \big\langle b_{l_1 m_1} b_{l_2 m_2} \big\rangle_0 \brac{ 1 + q^2 R^2 P_{l_1}  A_{l_1 m_1} }^{-1} \ .
\end{equation}
Upon comparing this with \eqref{eq: two point function momentum space exact} we find the prediction
\begin{equation} \label{eq: A prediction}
    A_{lm} = \frac{l(l+1)}{\pi} \ ,
\end{equation}
for the vacuum polarisation diagram's value.

Let us compute \eqref{eq: Alm def} and see whether it matches this prediction. First, we must be more precise about the integrals $\cal{A}_{lm\rho_1 \rho_2}$. This is most conveniently done by examining them for the four different combinations of $s_1$ and $s_2$, which gives
\begin{subequations}
\begin{align}
    \cal{A}^{+\sigma_1 + \sigma_2}_{lm L_1 M_1 L_2 M_2} &= a_{lm} c_{L_1 M_1} c_{L_2 M_2} F_{m M_1 M_2} \cal{I}_{lm L_1 M_1 L_2 M_2} \brac{1 - \sigma_1 \sigma_2 (-1)^{l + L_1 + L_2 - m - M_1 - M_2}} \ , \\
    \cal{A}^{-\sigma_1 - \sigma_2}_{lm L_1 M_1 L_2 M_2} &= - a_{lm} c_{L_1 M_1} c_{L_2 M_2} F^*_{m M_1 M_2} \cal{I}_{lm L_1 M_1 L_2 M_2} \brac{1 - \sigma_1 \sigma_2 (-1)^{l + L_1 + L_2 - m - M_1 - M_2}} \ , \\
    \cal{A}^{+\sigma_1 - \sigma_2}_{lm L_1 M_1 L_2 M_2} &= i a_{lm} c_{L_1 M_1} c_{L_2 M_2} G_{m M_1 M_2} \Big(\sigma_2 \cal{J}_{lm L_1 M_1 L_2 M_2} + \sigma_1 \cal{J}_{lm L_2 M_2 L_1 M_1}\Big) \ , \\
    \cal{A}^{- \sigma_1 + \sigma_2}_{lm L_1 M_1 L_2 M_2} &= - i a_{lm} c_{L_1 M_1} c_{L_2 M_2} G^*_{m M_1 M_2} \Big(\sigma_2 \cal{J}_{lm L_1 M_1 L_2 M_2} + \sigma_1 \cal{J}_{lm L_2 M_2 L_1 M_1}\Big) \ .
\end{align}
\end{subequations}
Here we are using the notation
\begin{subequations}
\begin{align} \label{eq: I integral}
    \cal{I}_{lm L_1 M_1 L_2 M_2} &= \int_0^{\pi} d\theta \, \sin\theta \, \Phi_{L_1 M_1}(\theta) \Phi_{L_2 M_2}(\theta) \, P^{|m|}_{l}(\cos\theta) \ , \\
    \cal{J}_{lm L_1 M_1 L_2 M_2} &= \int_0^{\pi} d\theta \, \sin\theta \, \Phi_{L_1 M_1}(\theta) \Psi_{L_2 M_2}(\theta) \, P^{|m|}_{l}(\cos\theta) \ , \\ 
    F_{mM_1 M_2} &= \int_0^{2\pi} d\phi \, f_{m}(\phi) e^{-i(M_1 - M_2)\phi} \ , \\
    G_{mM_1 M_2} &= \int_0^{2\pi} d\phi \, f_{m}(\phi) e^{-i(M_1 + M_2 + 1)\phi} \ ,
\end{align}
\end{subequations}
for the integrals over the sphere coordinates; see appendix \ref{sect: conventions} for the definitions of the functions on the sphere and normalisation constants used here. As only $\cal{A}_{lm\rho_1 \rho_2}$ depends on $s_i$, the sum over these variables gives
\begin{align} \nonumber
    \mathlarger{\sum}_{s_i} \cal{A}_{lm \rho_1 \rho_2} \cal{A}_{lm \rho_2 \rho_1} =  \ &\brac{a_{lm} c_{L_1 M_1} c_{L_2 M_2}}^2 \bigg[
    2 \brac{F_{m M_1 M_2}^2 + F_{m M_1 M_2}^{*\,2} }  \cal{I}_{lm L_1 M_1 L_2 M_2}^2 \\ \nonumber
    &\times\Big(1 -\sigma_1 \sigma_2 (-1)^{l + L_1 + L_2 - m - M_1 - M_2}\Big)
    + 2 \abs{G_{m M_1 M_2}}^2 \\
    &\times \Big(\sigma_2 \cal{J}_{lm L_1 M_1 L_2 M_2} + \sigma_1 \cal{J}_{lm L_2 M_2 L_1 M_1}\Big)^2 \bigg] \ .
\end{align}
As the diagram must be independent of $m$ we are free to evaluate it using a conveniently chosen value. We shall use $m=0$: as $f_0 = 1$ and $M_i \geq 0$, we find that
\begin{subequations}
\begin{align}
    F_{0 M_1 M_2} &= 2\pi \delta_{M_1, M_2} \ , \\
    G_{0 M_1 M_2} &= 0 \ ,
\end{align}
\end{subequations}
meaning the number of terms that must be computed in $A_{l0}$ is as small as possible. While it would be interesting to explicitly show that the result does not depend on $m$, we will not approach this question here. For $m=0$, the sum over $M_2$ can be performed to yield (after relabelling $M_1$ to $M$)
\begin{align} \nonumber
    A_{l 0} = -\mathlarger{\sum}_{L_i,\sigma_i} \mathlarger{\sum}_{M = 0}^{\min(L_1, L_2)} &\brac{4 \pi a_{l0} c_{L_1 M} c_{L_2 M} }^2 \Big(\sigma_1 (L_1 + 1) + \sigma_2 (L_2 + 1) \Big)^2 S_{L_1\sigma_1} S_{L_2 \sigma_2} \\
    &\times \cal{I}^2_{l0 L_1 M L_2 M} \brac{1 - \sigma_1 \sigma_2 (-1)^{l + L_1 + L_2}} \ .
\end{align}
We can sum over $\sigma_i$ by noting that this imposes the constraint
\begin{equation}
    \sigma_1 \sigma_2 = - (-1)^{l + L_1 + L_2} \equiv t_{l L_1 L_2} \ ,
\end{equation}
which has the solutions $\sigma_1 = 1, \sigma_2 = t_{l L_1 L_2}$ and $\sigma_1 = -1, \sigma_2 = -t_{l L_1 L_2}$. Using the explicit expressions for $a_{lm}$ and $c_{LM}$ given in appendix \ref{sect: conventions} and substituting in for $S_{L_i \sigma_i}$ we find
\begin{align} \nonumber
    A_{l0} = \frac{2l+1}{\pi} \mathlarger{\sum}_{L_i} \mathlarger{\sum}_{M = 0}^{\min(L_1, L_2)}& \frac{(L_1+M+1)! (L_2+M+1)! (L_1 - M)! (L_2 - M)! }{\brac{L_1!}^2 \brac{L_2 !}^2} \\
    & \times \frac{t_{l L_1 L_2} \brac{L_1 + 1 + t_{l L_1 L_2}\brac{L_2 + 1}}^2}{(L_1 + 1)(L_2 + 1)} \cal{I}^2_{l0 L_1 M L_2 M} \ .
\end{align}

\begin{figure} \hspace{-1.2cm}
    \includegraphics[width=1.15\linewidth]{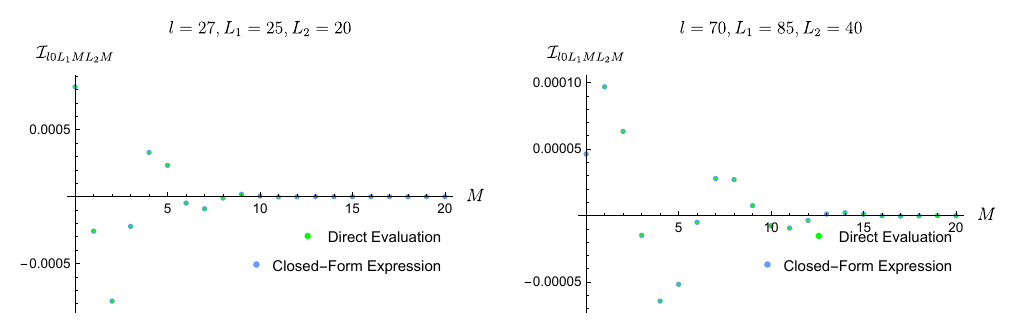}
    \caption{Comparison of direct evaluation of \eqref{eq: I integral} with the proposed expressions \eqref{eq: exact I even} and \eqref{eq: exact I odd} for varying values of $M$ when $l=27$, $L_1=25$, $L_2 = 20$, and $l=70$, $L_1 = 85$, $L_2=40$. In both cases the two exactly agree.}
    \label{fig: DE v CF}
\end{figure}

We will compute the sum by dividing it into
\begin{subequations}
\begin{align}
    A_{l0} &= \sum_{L_1=0}^{\infty} T_{l L_1} \ , \\ \nonumber
    T_{l L_1} &= \frac{2l+1}{\pi} \mathlarger{\sum}_{L_2\in \cal{S}_{l L_1}} \mathlarger{\sum}_{M = 0}^{\min(L_1, L_2)} \frac{(L_1+M+1)! (L_2+M+1)! (L_1 - M)! (L_2 - M)! }{\brac{L_1!}^2 \brac{L_2 !}^2} \\
    & \hspace{4.8cm} \times \frac{t_{l L_1 L_2} \brac{L_1 + 1 + t_{l L_1 L_2}\brac{L_2 + 1}}^2}{(L_1 + 1)(L_2 + 1)} \cal{I}^2_{l0 L_1 M L_2 M} \ ,
\end{align}
\end{subequations}
where the set $\cal{S}_{l L_1}$ is defined by
\begin{equation}
    \cal{S}_{l L_1} = \{ L_2 \mid \abs{L_1 - L_2} \leq l \leq L_1 + L_2 + 1  \} \ ,
\end{equation}
and denotes the values of $L_2$ for fixed $l$ and $L_1$ allowed by $SO(3)$-covariance. The sum defining $T_{l L_1}$ is then finite and can be explicitly computed if we know the value of $\cal{I}_{l0 L_1 M L_2 M}$. This is still a tall order, and to the best of our knowledge a direct computation of this integral for general parameters is not known (of course, direct evaluation for specified values of $l$, $L_i$, and $M$ is possible). However, by examining the value of the integral for various choices of parameters and finding sequences which match these values, starting with simple cases (such as when $l$ and $L_i$ are fixed but $M$ allowed to vary) and building up to include all parameters\footnote{And using a modicum of divine inspiration.}, we can propose a closed-form expression for the integral. It turns out that there are two cases we must consider, corresponding to whether the difference of $l$ and $\Delta = \abs{L_1 - L_2}$ is odd or even. For ease of notation we shall use $L$ to denote the smaller of $L_1$ and $L_2$. If $l-\Delta$ is even, we find
\begin{align} \nonumber
    \cal{I}_{l0 L_1 M L_2 M} = \, &\frac{(2L + \Delta - l+1)!! (l-\Delta-1)!! (l + \Delta-1)!!}{(2L + l + \Delta + 1)!! \brac{\frac{l-\Delta}{2}}! \brac{\frac{l+\Delta}{2}}! } \\ \label{eq: exact I even}
    &\times\frac{(L!)^2  \brac{L + \frac{l + \Delta }{2} +1}!}{(L-M)! (L+M+1)! \brac{L + \frac{\Delta - l}{2}}! (L+ \Delta + 1)} \sum_{n=0}^{(l-\Delta)/2} C_{l L \Delta M n} \ ,
\end{align}
where $C_{l L \Delta M n}$ is defined recursively by
\begin{equation}
    C_{l L \Delta M n} = \begin{cases}
        1 \ , & n=0 \\
        - \frac{(M+n)(M-n+1)(2n + l + \Delta-1) \brac{\frac{l - \Delta}{2}+1-n}}{(L-n+1)(L+\Delta +n +1)n (2 n+1)} C_{l L \Delta M (n-1)} \ , & 1 \leq n \leq L \\
        0 \ , & n> L
    \end{cases} \ .
\end{equation}
However, if $l-\Delta$ is odd we instead find
\begin{align} \nonumber
    \cal{I}_{l0 L_1 M L_2 M} = \, &\frac{(2M+1)(2L + \Delta - l)!! (l-\Delta)!! (l + \Delta)!!}{(2L + l + \Delta + 2)!! \brac{\frac{l-\Delta - 1}{2}}! \brac{\frac{l+\Delta - 1}{2}}! } \\ \label{eq: exact I odd}
    &\times\frac{(L!)^2  \brac{L + \frac{l + \Delta + 1}{2}}!}{(L-M)! (L+M+1)! \brac{L + \frac{\Delta - l + 1}{2}}! (L+ \Delta + 1)} \sum_{n=0}^{(l-\Delta-1)/2} C_{l L \Delta M n} \ ,
\end{align}
where $C_{l L \Delta M n}$ is now defined by
\begin{equation}
    C_{l L \Delta M n} = \begin{cases}
        1 \ , & n=0 \\
        - \frac{(M+n)(M-n+1)(2n + l + \Delta) \brac{\frac{l - \Delta + 1}{2}-n}}{(L-n+1)(L + \Delta +n +1)n (2 n+1)} C_{l L \Delta M (n-1)} \ , & 1 \leq n \leq L \\
        0 \ , & n> L
    \end{cases} \ .
\end{equation}
We should stress that we are not offering a proof of the results. While it would of course be of interest to prove these directly from the integral \eqref{eq: I integral}, we shall not pursue this here. However, one can check its validity for any values of $l$, $L_1$, and $L_2$; for instance, in figure \ref{fig: DE v CF} we show that this formula exactly agrees with direct evaluation of the integral \eqref{eq: I integral} for all allowed values of $M$ when $l=27$, $L_1 = 25$, and $L_2=20$, and for the first 20 values when $l=70$, $L_1=85$, and $L_2 = 40$\footnote{The values of $M$ past this point are practically indistinguishable from zero so we have dropped them, though they also agree with each other.}. In any theory on $S^2$ featuring cubic couplings between fermions and scalars we will see similar structure in the one-loop vacuum polarisation for the scalar (and ultimately in the partition function); the expression for $\cal{I}_{l0L_1 M L_2 M}$ therefore enables computations of such observables in a wide class of theories, which would be fascinating to study further.

With these results in hand, let us return to the calculation of $T_{l L_1}$. Considering this case-by-case, we experimentally find the somewhat miraculous result
\begin{equation} \label{eq: T values}
    T_{l L_1} = \begin{cases}
        \frac{L_1 + 1}{\pi} \ , \quad & L_1 \leq l - 1 \\
        0 \ , & L_1 \geq l 
    \end{cases}
\end{equation}
for $T_{l L_1}$. The computation of $A_{l0}$ then reduces to
\begin{align} \nonumber
    \sum_{L_1 = 0}^{\infty} \Tilde{T}_{l L_1} &=  \mathlarger{\sum}_{k=1}^l\, \frac{k}{\pi} \\ \label{eq: unregulated half of the answer}
    &= \frac{l(l+1)}{2\pi} \ ,
\end{align}
which is half the predicted value \eqref{eq: A prediction}. However, we saw this factor of two previously and therefore recognise it as the result of a regularization scheme that does not preserve gauge-invariance. The way this manifests in this calculation is that the sums performed to compute $T_{l L_1}$ are not absolutely convergent, meaning a naive summation should be thought of as a choice of regulator. It is somewhat novel to see the gauge anomaly appear this way, and points to the importance of proper regularisation of sums on the sphere in the same way as for integrals in the flat-space theory. As before, we can remedy the discrepancy using a Pauli-Villars regulator. Here we use a single commuting regulator field of mass $M = (\epsilon R)^{-1}$ whose effect is to shift the loop propagator to
\begin{equation}
    S_{L_1 \sigma_1} S_{L_2 \sigma_2} \to - \inv{\sigma_1 \sigma_2 (L_1 +1)(L_2 + 1)} - \inv{(i\sigma_1 (L_1 + 1) + \epsilon^{-1}) (i\sigma_2 (L_2 + 1) + \epsilon^{-1})} \ .
\end{equation}
The terms linear in $\sigma_i$ cancel in the sum over these variables, meaning we can effectively make the substitution
\begin{equation} \label{eq: regulated fermion loop ang mom}
    S_{L_1 \sigma_1} S_{L_2 \sigma_2} \to - \frac{t_{l L_1 L_2} (1 + \epsilon^2 ((L_1 + 1)^2 + (L_2 + 1)^2)}{(L_1 + 1)(L_2 + 1)(1 + \epsilon^2 (L_1 + 1)^2) (1 + \epsilon^2 (L_2 + 1)^2)}
\end{equation}
in the sum. If we  go through the same computation with the regularized propagator we find the same finite terms as before, as well as new terms which vanish in the  $\epsilon\to0$ limit. These are complicated, and an expansion in $\epsilon$ gives
\begin{align} \nonumber
    \Tilde{T}^{(\epsilon)}_{l L_1} = \,& \frac{l (l+1) (L_1 + 1) \epsilon^2}{\pi} \bigg[
    1 - \epsilon^2 \Big(2 (L_1 + 1)^2 - 1 + l(l+1) \Big) + \epsilon^4 \Big( 3 (L_1 + 1)^4 \\ \label{eq: T tilde exp}
    &- 5 (L_1 + 1)^2 + 3 + l(l+1)( 4 (L_1 + 1)^2 - 3) + l^2 (l+1)^2 \Big) + O(\epsilon^6) \bigg] \ ,
\end{align}
for the first three terms. While each term is individually divergent when summed over $L_1$, one expects that the resummation of the bracketed terms leads to a convergent sum. However, as we can only compute the terms in the expansion order-by-order it is difficult to see what the full resummed function is; at first glance, it therefore appears our inability to write these terms in a closed-form expression rules out any computation of the sum of $\Tilde{T}^{(\epsilon)}_{l L_1}$. Luckily for us, this is not the case if one is only interested in the $\epsilon \to 0$ limit of this sum for the following reason. From the relation between $\epsilon$ and $k=L_1 + 1$ in \eqref{eq: regulated fermion loop ang mom} and the overall factors of each in \eqref{eq: T tilde exp}, we see that the new terms take the form
\begin{align} \nonumber
    \Tilde{T}^{(\epsilon)}_{l L_1} &= k \epsilon^2  \sum_{n=0}^{\infty} \epsilon^{2n} \sum_{m=0}^{\infty} a_{m,n,l} (k\epsilon)^{2m} \\
    &= k \epsilon^2 \sum_{n=0}^{\infty} \epsilon^{2n} f_{n,l}(k\epsilon) \ ,
\end{align}
where we have introduced the resummed functions $f_{n,l}$ whose expansions at small argument have the coefficients $a_{m,n,l}$\footnote{In essence, we are simply undoing the expansion in $\epsilon$ performed to reach \eqref{eq: T tilde exp}.}. The sum over $k$ then gives
\begin{equation} \label{eq: sum of T tilde}
    \sum_{k=1}^{\infty} \Tilde{T}^{(\epsilon)}_{l L_1} = \sum_{n=0}^{\infty} \epsilon^{2n} \sum_{k=1}^{\infty} k \epsilon^2 f_{n,l}(k\epsilon) \ ,
\end{equation}
where we have presumed each sum is sufficiently convergent for there to be no issue interchanging them (while we believe this to be the case, we should stress that this is an assumption we are making on the functions $f_{n,l}$). However, we see that as we take $\epsilon\to 0$ the sum over $k$ becomes the integral
\begin{equation} \label{eq: sum to integral}
    \sum_{k=1}^{\infty} k \epsilon^2 f_{n,l}(k\epsilon) \to \int_0^{\infty} dx \, x f_{n,l}(x) \ ,
\end{equation}
for each value of $n$, which (again, provided the functions $f_{n,l}$ are sufficiently nice for these integrals to exist) is finite. In the $\epsilon\to0$ limit the only term with a non-vanishing contribution to \eqref{eq: sum of T tilde} is then when $n=0$.

\begin{figure}[htp] 
\begin{center}
\begin{subfigure}{\textwidth}
\begin{center}
\includegraphics[width=0.8\textwidth]{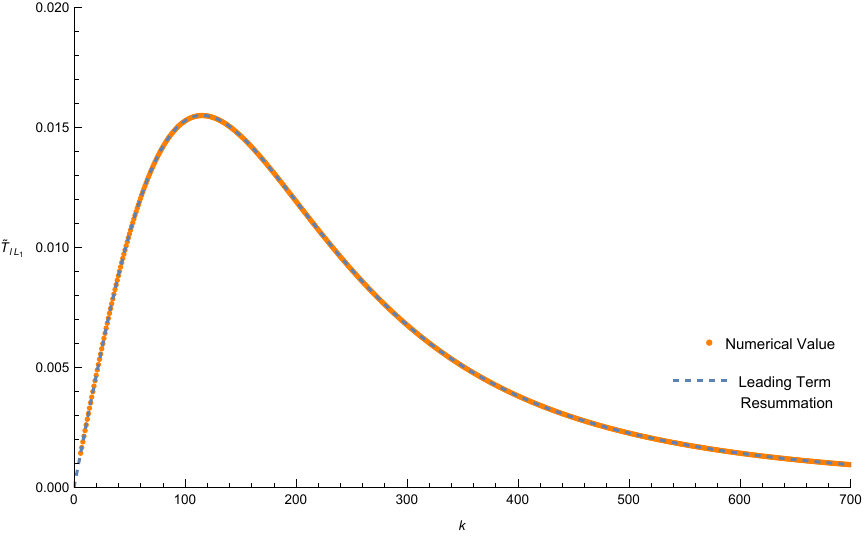}
\end{center}
\end{subfigure}

\begin{subfigure}{\textwidth}
\begin{center}
\includegraphics[width=0.8\textwidth]{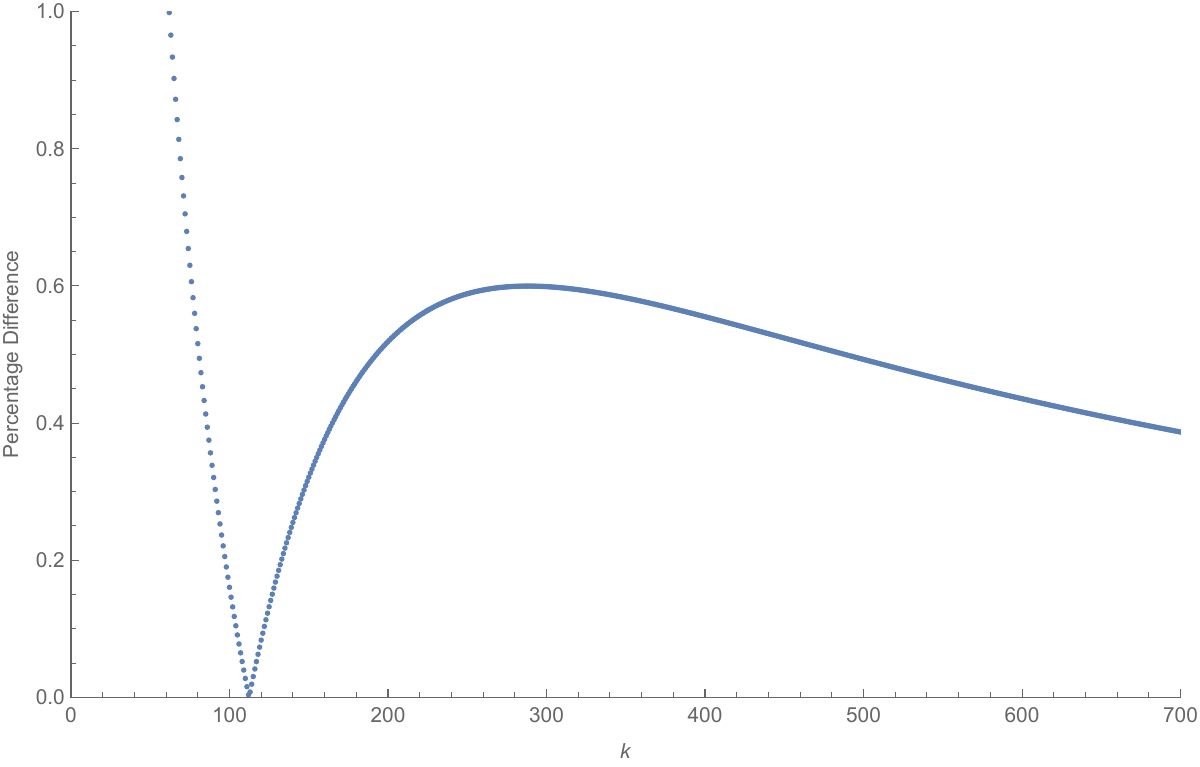}
\end{center}
\end{subfigure}

\caption{Numerical and proposed leading-order resummed values of $\Tilde{T}_{l L_1}$ and the percentage difference $100\times |\Tilde{T}^{(\text{num.})}_{l L_1} -\Tilde{T}^{(\text{resum.})}_{l L_1} |\,/\,\Tilde{T}^{(\text{num.})}_{l L_1}$ between the two for $l=5$ and $\epsilon=0.005$ up to $k=700$.} \label{fig: test of leading resummation}
\end{center}
\end{figure}

The argument above means we can restrict our attention to
\begin{equation}
   \Tilde{T}^{(\epsilon)}_{l L_1} = \frac{l(l+1) k \epsilon^2}{\pi} \brac{1 - 2 \epsilon^2 k^2 + 3 \epsilon^4 k^4 + O(\epsilon^6)} + ... \ .
\end{equation}
From this simpler expression we can propose the resummed expression
\begin{equation}
    \Tilde{T}^{(\epsilon)}_{l L_1} = \frac{l (l+1) k \epsilon^2}{\pi(1 + k^2 \epsilon^2)^2} + ...  
\end{equation}
for the terms of interest. In figure \ref{fig: test of leading resummation} we show numerically that this is an exceptionally good approximation to the exact value of the terms, and will assume from here onwards that this captures their large-$k$, or equivalently small-$\epsilon$, behaviour. Helpfully, the sum of these terms takes the closed form
\begin{equation}
    \sum_{k=1}^{\infty} \frac{l (l+1) k \epsilon^2}{\pi (1 + k^2 \epsilon^2)^2} = - \frac{il(l+1)}{4\pi\epsilon} \brac{ \psi^{(1)}(1 - i \epsilon^{-1}) - \psi^{(1)}(1 + i \epsilon^{-1}) } \ ,
\end{equation}
where $\psi^{(1)}(z)$ is the polygamma function of order 1. Taking the $\epsilon\to0$ limit of this gives 
\begin{equation}
    \lim_{\epsilon\to0} \sum_{L_1 = 0}^{\infty} \Tilde{T}^{(\epsilon)}_{l L_1} = \frac{l (l+1)}{2\pi} \ .
\end{equation}
Of course, in the $\epsilon\to 0$ limit we find the same result from \eqref{eq: sum to integral} using
\begin{equation}
    \int_0^{\infty}  \frac{dx\,x}{(1+x^2)^2} = \inv{2} \ .
\end{equation}
Combining this with the value \eqref{eq: unregulated half of the answer} arising from the finite terms gives
\begin{align} \nonumber
    A_{l0} &= \lim_{\epsilon\to 0 } \sum_{L_1 = 0}^{\infty} \brac{ T_{l L_1} + \Tilde{T}^{(\epsilon)}_{l L_1}} \\
    &= \frac{l(l+1)}{\pi} \ ,
\end{align}
which now exactly matches the predicted value.


\subsection{The Partition Function to All Orders in Perturbation Theory}

We can use this result to perturbatively compute the theory's partition function. Let us first start with the two-loop contribution; the diagram \eqref{eq: two-loop partition function diagram} in momentum space is
\begin{align} \nonumber
\begin{gathered}
    \includegraphics[width=0.165\textwidth]{FeynmanDiagramPartitionFunction.pdf}
\end{gathered} 
 &= - \frac{q^2 R^2}{2} \sum_{l,m} P_l A_{lm} \\[-0.3em] \label{eq: angular momentum partition function}
 & = - \frac{q^2 R^2}{2\pi} \mathlarger{\sum}_{l=1}^{\infty} \frac{2l + 1}{l(l+1) }  \ ,
\end{align}
where we use \eqref{eq: A prediction} and sum over $m$ to reach the second equality. This is clearly a divergent sum, as expected from the logarithmic divergence present in the exact solution. We can regulate this in many ways, but a convenient choice is to introduce a Pauli-Villars partner of the prepotential field with a mass $\Lambda^{-1} = \epsilon R$. The effect of this is to alter the propagator to 
\begin{align} \nonumber
    P_l \to \Tilde{P}_l &= \inv{l^2(l+1)^2} - \inv{l(l+1) \brac{l(l+1) + \Lambda^2 R^2}} \\
    &\equiv \inv{l^2(l+1)^2 \brac{1 + \epsilon^2 l(l+1)}} \ .
\end{align}
Using this, \eqref{eq: angular momentum partition function} becomes
\begin{align}
    - \frac{q^2 R^2}{2} \mathlarger{\sum}_{l,m} \Tilde{P}_l A_{lm} = - \frac{q^2 R^2}{2\pi} \mathlarger{\sum}_{l=1}^{\infty} \frac{2l+1}{l(l+1)\brac{1 + \epsilon^2 l(l+1)}} \ .
\end{align}
This sum can be explicitly performed, and we find
\begin{equation}
    \ln\brac{\frac{\cal{Z}_q}{\cal{Z}_0}}\bigg\rvert_{q^2 R^2} = \frac{q^2 R^2}{2\pi} \brac{ 1- 2 \gamma - \psi\left(\frac{3}{2} -\frac{  \sqrt{\epsilon ^2-4}}{2 \epsilon}\right) - \psi\left(\frac{3}{2} + \frac{  \sqrt{\epsilon ^2-4}}{2 \epsilon}\right)} \ ,
\end{equation}
where $\psi$ is the digamma function. Expanding this for small $\epsilon$ gives
\begin{equation}
    \ln\brac{\frac{\cal{Z}_q}{\cal{Z}_0}}\bigg\rvert_{q^2 R^2} = \frac{q^2 R^2 \brac{2\ln \epsilon  + 1 -2 \gamma}}{2\pi}  + O(\epsilon^2) \ ,
\end{equation}
which matches the term \eqref{eq: partition function exact result} in the expansion of the exact result. In fact, as integrating out the fermion is one-loop exact the only diagrams that contribute to the partition function are photon loops with repeated insertions of the vacuum polarisation diagram. As this takes an incredibly simple form we can easily compute further corrections. For instance, the next term in the expansion comes from the diagram
\begin{center}
    \includegraphics[width=0.37\textwidth]{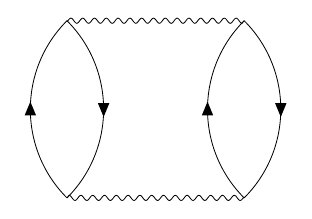} 
\end{center}
and so we find
\begin{align} \nonumber
    \ln\brac{\frac{\cal{Z}_q}{\cal{Z}_0}}\bigg\rvert_{q^{4} R^{4}} &= \frac{q^4 R^4}{4\pi^2} \mathlarger{\sum}_{l=1}^{\infty} \frac{2l+1}{l^2 (l+1)^2} \\
    &= \frac{q^4 R^4}{4\pi^2} \ ,
\end{align}
which again matches the $O(q^4 R^4)$ term in the expansion of the exact solution. More generally, the $O(q^{2n} R^{2n})$ term is
\begin{align} \nonumber
    \ln\brac{\frac{\cal{Z}_q}{\cal{Z}_0}}\bigg\rvert_{q^{2n} R^{2n}} &= (-1)^n \frac{q^{2n} R^{2n}}{(2\pi)^n n!} \mathlarger{\sum}_{l_i,m_i} l_1 (l_1 + 1) ... l_n (l_n + 1) \langle b_{l_1 m_1}^2 ... b_{l_n m_n}^2 \rangle_0^{(conn.)} \\
    &= (-1)^n \frac{q^{2n} R^{2n} }{2n \pi^n } \mathlarger{\sum}_{l=1}^{\infty} \frac{2l+1}{l^n(l+1)^n} \ ,
\end{align}
after expanding out the connected correlation function using Wick contractions. Assuming $n\geq2$, we can decompose the summand into partial fractions to find
\begin{equation}
    \frac{2l + 1}{l^n (l+1)^n} = \sum_{k=0}^{n-2} a_{k}^{(n)} \brac{
    \frac{(-1)^k}{l^{n-k}} - \frac{(-1)^n}{(l+1)^{n-k}} } \ ,
\end{equation}
where the coefficient $a_k^{(n)}$ is defined recursively by
\begin{subequations}
\begin{align}
    a_0^{(n)} &= 1 \ , \\
    a_k^{(n)} &= \begin{cases}
        a_{k-1}^{(n)} + a^{(n-1)}_k \ , & 0<k\leq n-2 \\
        0 \ , & k > n-2
    \end{cases} \ .
\end{align}
\end{subequations}
The sum over $l$ can then be performed, leading to the result
\begin{equation}
    \mathlarger{\sum}_{l=1}^{\infty} \frac{2l+1}{l^n (l+1)^{n}} = (-1)^n \mathlarger{\sum}_{k=0}^{n-2} a_k^{(n)} \bigg[
    1 - \brac{1 - (-1)^{n-k}} \zeta(n-k) \bigg] \ ,
\end{equation}
and so the $O(q^{2n} R^{2n})$ term in the perturbative expansion of the sphere partition function is
\begin{equation}
    \ln\brac{\frac{\cal{Z}_q}{\cal{Z}_0}}\bigg\rvert_{q^{2n} R^{2n}} = \frac{q^{2n} R^{2n}}{2n \pi^n } \mathlarger{\sum}_{k=0}^{n-2} a_k^{(n)} \bigg[ 1 - \brac{1 - (-1)^{n-k}} \zeta(n-k) \bigg] \ ,
\end{equation}
which can be checked to agree order-by-order with the expansion of the exact solution. For instance, when $n=7$ we have $a^{(7)}_k = 1,5,14,28,42,42,0,0,...\,$, and so 
\begin{align} \nonumber
    \ln\brac{\frac{\cal{Z}_q}{\cal{Z}_0}}\bigg\rvert_{q^{14} R^{14}} &= \inv{14} \brac{\frac{q^2 R^2}{\pi}}^7 \Bigg[
    \sum^5_{k=0} a^{(7)}_k - 2 \brac{ a^{(7)}_0 \zeta(7) + a^{(7)}_2 \zeta(5) + a^{(7)}_4 \zeta(3) } \Bigg] \\
    &= - \inv{7} \brac{\frac{q^2 R^2}{\pi}}^7 \bigg(
    \zeta(7) + 14 \zeta(5) + 42\zeta(3) - 66 \bigg) \ ,
\end{align}
which is the result given in appendix D of \cite{Anninos:2024fty}.

\section{Conclusion} \label{sect: conclusion}

In this note we have examined the perturbative structure of the Schwinger model on $S^2$ through computations of the one-loop correction to the photon propagator and the partition function. We did this using both a numerical position-space computation in stereographic coordinates and a semi-analytic calculation in momentum-space. In both cases we found agreement with the exact solution of \cite{Jayewardena:1988td, Anninos:2024fty} once a gauge-invariant regularisation scheme was used. In the momentum-space approach we saw that the effect of the gauge anomaly was realised as a conditionally convergent sum, pointing to the importance of proper regularisation for gauge theories on $S^2$. The two methods were seen to have differing strengths; the position-space calculation, though simpler, only yielded numerical results, whereas the momentum-space approach relied on a somewhat convoluted determination of the interaction vertex but yielded the exact result for each observable.

In our momentum-space calculation we restricted to the case $m=0$ for simplicity, which led us to conjecture the values \eqref{eq: exact I even} and \eqref{eq: exact I odd} for $\cal{I}$ in the interaction vertex. It would be of interest to extend this to find the values of $\cal{I}$ and $\cal{J}$ for any $m$ in order to completely characterise the momentum-space vertex. This would, for example, allow us to show explicitly that the vacuum polarisation diagram is $m$-independent. Additionally, a complete result for the interaction vertex would also allow for computations of higher-loop corrections in the Schwinger model; predictions for these can be found through an expansion of the known exact results, and it would be instructive to check that the two match. 

More generally, expressions for $\cal{I}$ and $\cal{J}$ would allow for computations of observables in any QFT on the sphere featuring a cubic coupling between a scalar field and fermionic fields; while we focused here on the Schwinger model, a similar momentum-space expansion structure will occur in any such theory. Since our understanding of loop effects in higher-dimensional de Sitter QFTs are not well-developed, it would be worthwhile to understand quantum corrections in the two-dimensional case to better probe the role and validity of perturbation theory for QFTs in expanding cosmologies. While we did not give a complete expression for $\cal{I}$ in this work, our restricted formula is still sufficient to compute the contribution of fermions to the partition function of such theories. It would be intriguing to adapt the computation performed here to other theories of interest for which an exact result is not known, such as the $dS_2$ supergravity theories proposed in \cite{Anninos:2023exn}.

\section*{Acknowledgments}

I thank Dionysios Anninos for the suggestion of this topic and many helpful discussions during the completion of this work, Omar Shahpo for his help with performing the numerical computations in section \ref{sect: stereo pert}, and Faisal Karimi and Alan Rios Fukelman for further discussions. I was partially supported by the STFC studentship ST/W507556/1 during the completion of this work.

\appendix

\section{Conventions} \label{sect: conventions}

In this appendix we will review the conventions used throughout this paper, starting with those of the fermions. The curved-space gamma matrices are given by
\begin{equation}
    \gamma^{\mu} = e^{\mu}_a \pmb{\sigma}^a \ ,
\end{equation}
where in stereographic coordinates we use the orthonormal basis
\begin{equation}
    e^{\mu}_a(\bf{x}) = \Omega \brac{\frac{\partial}{\partial \bf{x}^a}}^{\mu} \ ,
\end{equation}
and the matrices $\pmb{\sigma}^a$ obey the anticommutation relations
\begin{equation}
    \{\pmb{\sigma}^a , \pmb{\sigma}^b\} = 2 \delta^{ab} \ .
\end{equation}
When an explicit realisation of this is needed we will use the Pauli matrix representation
\begin{subequations}
\begin{align}
    \pmb{\sigma}^1 = \begin{pmatrix}
        0 & 1 \\
        1 & 0
    \end{pmatrix} \ , \\
    \pmb{\sigma}^2 = \begin{pmatrix}
        0 & -i \\
        i & 0
    \end{pmatrix} \ .
\end{align}
\end{subequations}
Using these we can construct
\begin{equation}
    \gamma_* = - i \pmb{\sigma}^1 \pmb{\sigma}^2 \ ,
\end{equation}
which as usual anticommutes with both $\pmb{\sigma}^1$ and $\pmb{\sigma}^2$. In the Pauli matrix representation this is just
\begin{equation}
    \gamma_* \equiv \pmb{\sigma}^3 = \begin{pmatrix}
        1 & 0 \\
        0 & -1
    \end{pmatrix} \ .
\end{equation}
The covariant derivative of our spinor field $\psi$ is defined as
\begin{equation}
    \nabla_{\mu} \psi = \partial_{\mu} \psi + \inv{4} \omega_{\mu}^{ab} \pmb{\sigma}_{ab} \psi \ ,
\end{equation}
with $\omega_{\mu}$ the spin connection of $e^{\mu}_a$.

We will also require expansions of our fields on the sphere. For the scalar field $\beta$, the expansion is most naturally defined in terms of real spherical harmonics $\cal{Y}_{lm}$, where as usual $l\in\bb{Z}_{\geq0}$ and $m=-l,-1+1, ... \,, l$. In terms of the more familiar complex spherical harmonics $Y_l^m$ these are
\begin{equation}
    \cal{Y}_{lm} = \begin{cases}
        \frac{i}{\sqrt{2}} \brac{ Y^{-|m|}_l - (-1)^m Y^{|m|}_l} \ , & m<0  \\
        Y^0_l \ , & m=0  \\
        \inv{\sqrt{2}} \brac{ Y^{-|m|}_l + (-1)^m Y^{|m|}_l} \ , & m>0 
    \end{cases} \ ,
\end{equation}
and obey the equation
\begin{equation}
    \nabla^2 \cal{Y}_{lm} = - \frac{l(l+1)}{R^2} \cal{Y}_{lm} \ .
\end{equation}
Explicitly, we have
\begin{equation}
    \cal{Y}_{lm}(\theta,\phi) = a_{lm} f_{m}(\phi) P^{|m|}_l(\cos\theta) \ ,
\end{equation}
in terms of the associated Legendre polynomials $P^{|m|}_l(\cos\theta)$, with
\begin{equation}
    a_{lm} = \sqrt{\frac{(2l+1)(l-|m|)!}{4\pi (l+|m|)!}} \ ,
\end{equation}
and
\begin{equation}
    f_{m}(\phi) = \begin{cases}
        (-1)^m \sqrt{2} \sin\brac{|m|\phi} \ , & m<0  \\
        1 \ , & m=0  \\
        (-1)^m \sqrt{2} \cos\brac{|m|\phi} \ , & m>0 
    \end{cases} \ .
\end{equation}
Similarly, the fermionic fields are expanded using the eigenspinors $\psi^{s\sigma}_{LM}$ of the Dirac equation,
\begin{equation}
    \gamma^{\mu} \nabla_{\mu} \psi^{s \sigma}_{LM} = i \sigma \brac{L+1} \psi^{s \sigma}_{LM} \ ,
\end{equation}
with $L\in \bb{Z}_{\geq 0}$, $M = 0 , 1 , ... \, , L$, and $s,\sigma = \pm1$. The explicit forms of the spinors are \cite{Camporesi:1995fb}
\begin{subequations}
\begin{align}
    \psi^{+ \sigma}_{LM} &= c_{LM} e^{i (M + 1/2) \phi} \begin{pmatrix}
        \Phi_{LM}(\theta) \\
        i \sigma \Psi_{LM}(\theta)
    \end{pmatrix} \ , \\
    \psi^{- \sigma}_{LM} &= c_{LM} e^{-i (M + 1/2) \phi} \begin{pmatrix}
        i \sigma \Psi_{LM}(\theta) \\
        \Phi_{LM}(\theta)
    \end{pmatrix} \ ,
\end{align}
\end{subequations}
where we use the functions
\begin{subequations}
\begin{align}
    \Phi_{LM}(\theta) &= \cos^{M+1}\brac{\frac{\theta}{2}} \sin^{M}\brac{\frac{\theta}{2}} P^{(M,M+1)}_{L-M}(\cos\theta) \ , \\
    \Psi_{LM}(\theta) &= \brac{-1}^{L-M} \Phi_{LM}(\pi - \theta) \ ,
\end{align}
\end{subequations}
defined in terms of the Jacobi polynomials $P^{(M,M+1)}_{L-M}(\cos\theta)$, and use the notation
\begin{equation}
    c_{LM} = \sqrt{\frac{(L+M+1)!(L-M)!}{4\pi L!^2}} \ .
\end{equation}

\section{The Gauge Anomaly and the Partition Function} \label{sect: gauge anomaly}

In this appendix we shall review gauge anomalies in the Schwinger model and how they affect the computation of the partition function. To do this, let us first discuss the case of a free massless fermion coupled to a background $U(1)$ gauge field,
\begin{subequations}
\begin{align}
    \cal{Z}_F[\cal{A}] &= \bigintsss D\bpsi D\psi \, e^{-S_F[\bpsi,\psi,\cal{A}]} \ , \\
    S_F[\bpsi,\psi,\cal{A}] &= \bigintsss d^2x \, \sqrt{g} \, \bpsi \gamma^{\mu} \brac{\nabla_{\mu} + i qR \cal{A}_{\mu}} \psi \ .
\end{align}
\end{subequations}
We shall be interested in the one-point function of the $U(1)$ current which couples to the gauge field; a perturbative expansion of
\begin{align}
    \langle J^{\mu}(x) \rangle_{\cal{A}} = \inv{\cal{Z}[\cal{A}]} \bigintsss D\bpsi D\psi \, \brac{\bpsi \gamma^{\mu} \psi}(x) \, e^{-S[\bpsi,\psi,\cal{A}]}
\end{align}
in $qR$, using stereographic coordinates and the conformal rescaling \eqref{eq: flat fermion} of the fermions, yields
\begin{align} \nonumber
    \big\langle J^{a}(\bf{x}) \big\rangle_{\cal{A}} = \inv{\Omega(\bf{x})^2} \bigg[& 
    \tr\brac{S_0(\bf{x},\bf{x}) \pmb{\sigma}^a} + i qR \int d^2 \bf{y} \, \cal{A}_b(\bf{y}) \tr \brac{\pmb{\sigma}^a S_0(\bf{x},\bf{y}) \pmb{\sigma}^b S_0(\bf{y},\bf{x}) } \bigg] \\ \label{eq: J expectation}
    &+ O(q^2R^2) \ .
\end{align}
In this coordinate system the divergence of the current is
\begin{align}
    \big\langle \nabla_{\mu} J^{\mu}(x) \big\rangle_{\cal{A}} = \inv{\Omega(\bf{x})^2} \partial_a \bigg(
    \Omega(\bf{x})^2 \big\langle J^a(\bf{x}) \big\rangle_{\cal{A}} \bigg) \ ,
\end{align}
from which we see that the explicit factor of $\Omega^2$ within the derivative cancels the factor in \eqref{eq: J expectation}. Assuming the regularisation scheme we use is such that $\tr\brac{\sigma^a S_0(\bf{x},\bf{x})} =0$, which shall be true in all cases we consider, we see that the first-order contribution to the current's divergence is
\begin{align} \label{eq: current conservation anomaly}
    \Omega(\bf{x})^2 \big\langle \nabla_{\mu} J^{\mu}(x) \big\rangle_{\cal{A}}\Big\rvert_{qR} = i q R \, \partial_a \bigg(
    \int d^2 \bf{y} \, \cal{A}_b(\bf{y}) \tr \brac{\pmb{\sigma}^a S_0(\bf{x},\bf{y}) \pmb{\sigma}^b S_0(\bf{y},\bf{x}) }
    \bigg) \ .
\end{align}
The right of this equation is exactly what one would find in flat space, and so a regularisation of massless fermions on $S^2$ is gauge-invariant only if it implies a gauge-invariant regularisation of the rescaled fields \eqref{eq: flat fermion} in flat space.

We can now go further and see what effect this has on the partition function of the theory. If we work in the sector with vanishing instanton number the background gauge field can be decomposed into \cite{Jayewardena:1988td}
\begin{equation} \label{eq: A decomp}
    \cal{A} = d \alpha + *d\beta \ ,
\end{equation}
with $\alpha$ and $\beta$ scalar functions on $S^2$. The coupling of $\cal{A}$ to the fermion's $U(1)$ current can then be rewritten as
\begin{align}
    \int d^2x \sqrt{g} \, A_{\mu} J^{\mu} = - \int d^2x \sqrt{g} \brac{
    \alpha \, \nabla_{\mu} J^{\mu} - \beta \, \nabla_{\mu} J^{\mu}_A} \ ,
\end{align}
where $J_A$ is given by
\begin{align} \nonumber
    J_A^{\mu} &= \varepsilon^{\mn} J_{\nu} \\
    &\equiv i \bpsi \gamma^{\mu} \gamma_* \psi \ ,
\end{align}
which we recognise as the usual axial current. The dependence of the partition function on the scalar fields determines whether current conservation is preserved quantum-mechanically, as
\begin{subequations} \label{eq: conservation as response}
\begin{align}
    i q R \langle \nabla_{\mu} J^{\mu}(x) \rangle_{\cal{A}} &= \inv{\sqrt{g(x)}} \frac{\delta \cal{Z}}{\delta \alpha(x)} \ , \\
    -i q R \langle \nabla_{\mu} J_A^{\mu}(x) \rangle_{\cal{A}} &= \inv{\sqrt{g(x)}} \frac{\delta \cal{Z}}{\delta \beta(x)} \ .
\end{align}
\end{subequations}
Since both currents are classically conserved, the partition function can only be modified by an anomaly action involving $\alpha$ and $\beta$ arising from non-invariance of the measure \cite{Fujikawa:1979ay} under vector and axial $U(1)$ transformations which can be shown to be local. The action is constrained to vanish when $q=0$ and can only contain derivative terms in the fields, and so by dimensional analysis must take the form
\begin{subequations}
\begin{align} \label{eq: fermion partition function}
    \cal{Z}_F[\alpha,\beta] &= e^{-S_{anom.}[\alpha,\beta]} \cal{Z}_F[0] \ , \\ \nonumber
    S_{anom.}[\alpha,\beta] &= -\frac{q^2 R^2}{2\pi} \int d^2 x \,\sqrt{g} \brac{
    a \,\nabla^{\mu} \alpha \nabla_{\nu} \alpha - b \, \nabla^{\mu} \beta \nabla_{\nu} \beta } \\
    &\equiv S_{\alpha}[\alpha] + S_{\beta}[\beta] \ .
\end{align}
\end{subequations}
Combining this with \eqref{eq: conservation as response} we find that
\begin{subequations}
\begin{align}
    \langle \nabla_{\mu} J^{\mu}(x) \rangle_{\cal{A}} &= i \frac{a \, q R}{\pi} \nabla^2 \alpha(x) \ , \\
    \langle \nabla_{\mu} J_A^{\mu}(x) \rangle_{\cal{A}} &= i \frac{b \, q R}{\pi} \,\nabla^2 \beta(x) \ ,
\end{align}
\end{subequations}
and we can therefore find $a$ from the one-loop integral \eqref{eq: current conservation anomaly}. It is well-known from flat-space calculations that with this normalisation the two coefficients are related by\footnote{See, for example, \cite{Peskin:1995ev} for a textbook discussion of the regulator dependence of the two expectation values, with the key point being that the sum of the two coefficients is determined by a finite regulator-independent integral.}
\begin{equation}
    a + b = 1 \ ,
\end{equation}
so this is sufficient to characterise the anomaly structure.

Suppose we now try to perturbatively compute the observables we are interested in, namely $\ln\cal{Z}_q$ or the prepotential two-point function, using a regulator that induces a gauge-anomaly (i.e. $b\neq1$): what is the effect of the anomaly on our calculation? In other words, how do we see that our regularisation scheme is anomalous directly within perturbation theory? Let us focus on the two-point function for definiteness. Recall that the Schwinger model partition function \eqref{eq: partition function} is
\begin{equation}
    \cal{Z}_q = \int \frac{D A}{\text{vol}(\cal{G})} \, e^{-S_A[A]} \, \cal{Z}_F[A] \ .
\end{equation}
To proceed, we shall change integration variables from $A\to (\alpha,\beta)$ defined in \eqref{eq: A decomp}. As $\alpha$ is pure gauge the action $S_A[A]$ is only a functional of $\beta$, and using \eqref{eq: F to phi} is explicitly given by
\begin{equation}
    S_A[\beta] = \frac{R^2}{2} \int d^2 x \sqrt{g} \, \nabla^2 \beta \nabla^2 \beta \ .
\end{equation}
Using \eqref{eq: fermion partition function} and combining the $\beta$-dependent terms in the action, we can rewrite $\cal{Z}_q$ in the form
\begin{align} \label{eq: partition function for any anomaly}
    \cal{Z}_q &= \bigintssss D\beta \, \cal{N}  e^{-\Tilde{S}_{\beta}[\beta]} \ , \\
    \Tilde{S}_{\beta}[\beta] &= \frac{R^2}{2} \bigintssss d^2 x \sqrt{g} \, \beta \brac{\nabla^2 - \frac{q^2 b}{\pi}} \nabla^2 \beta \ , \\
    \cal{N} &= \cal{Z}_F[0] \bigintsss \frac{D \alpha}{\text{vol}(\cal{G})} J_{\alpha,\beta} e^{-S_{\alpha}[\alpha]} \ ,
\end{align}
where $J_{\alpha,\beta}$ is the ($\beta$-independent) Jacobian associated with the field transformation from $A$ to $(\alpha, \beta)$. As $\cal{N}$ is completely $\beta$-independent, it cancels within the prepotential two-point function
\begin{align}
    \langle \beta(x) \beta(y) \rangle_q = \cal{Z}_q^{-1} \bigintssss D \beta \cal{N} \, \beta(x) \beta(y) e^{-\Tilde{S}_{\beta}[\beta]} \ ,
\end{align}
which can therefore be computed by the action $\Tilde{S}_{\beta}[\beta]$. The only difference between this action and that obtained for the non-anomalous regularisation is that the mass of $\beta$ for each choice of regulator differs by a term proportional to the anomaly coefficient $a$,
\begin{equation}
    m^2_{anom.} = m^2 - \frac{q^2 a}{\pi} \ ,
\end{equation}
with $m^2 = q^2/\pi$ the mass obtained for a regulator which respects gauge-invariance. Practically speaking, when we expand the two-point in powers of $q$ for different regulators we will find a multiplicative factor relating terms at each order. If we use the notation
\begin{equation}
    \langle \beta(x) \beta(y)\rangle_q^{(b)} = \sum_{n=0}^{\infty} q^{2n} G^{(b)}_n(x,y) \ ,
\end{equation}
for the expansion of $\langle \beta(x) \beta(y)\rangle_q$ for a regulator with anomaly $b$, we find that
\begin{equation}
    G^{(b)}_{n}(x,y) = b^{n} G_n(x,y) \ ,
\end{equation}
where $G_n(x,y)$ is the non-anomalous result. Similar considerations show that the same conclusion also holds for the perturbative expansion of $\ln\cal{Z}_q$, with terms computed using an anomalous regulator differing from the non-anomalous terms by a factor of $b^n$ at order $q^{2n}$.

With this in mind, let us numerically compute the anomalies of the two regulators used in section \ref{sect: stereo pert}. Explicitly, these are given by the (unregulated) integrals
\begin{subequations} \label{eq: anomaly numerics}
\begin{align}
    \frac{a}{\pi} \partial_a \partial_a \alpha(\bf{x}) &= \partial_a \brac{
    \int d^2 \bf{y} \, \cal{A}_b(\bf{y}) \tr \brac{ \pmb{\sigma}^a S_0(\bf{x},\bf{y}) \pmb{\sigma}^b S_0(\bf{y},\bf{x})} } \ , \\
    \frac{b}{\pi} \partial_a \partial_a \beta (\bf{x}) &= i \partial_a \brac{
    \int d^2 \bf{y} \, \cal{A}_b(\bf{y}) \tr \brac{ \pmb{\sigma}^a \pmb{\sigma}^3 S_0(\bf{x},\bf{y}) \pmb{\sigma}^b S_0(\bf{y},\bf{x})} } \ .
\end{align}
\end{subequations}
\begin{figure}
    \centering
    \includegraphics[width=0.8\linewidth]{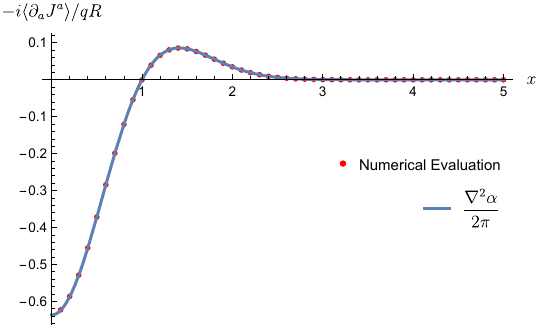}
    \caption{Numerical evaluation of the one-loop contribution to the divergence of $J^a$ in the minimum length regularization \eqref{eq: min length reg} with $\epsilon= 0.01$; the exact result for the anomaly coefficient $a=1/2$ is given for comparison.}
    \label{fig: min length anomaly}
\end{figure}
For convenience let us take
\begin{equation}
    \alpha = \beta = \exp(- \bf{x}\cdot \bf{x}) \ ,
\end{equation}
and evaluate both sides of \eqref{eq: anomaly numerics} along $\bf{x} = (x,0)$. Let's start with the cut-off lengthscale regularization \eqref{eq: min length reg}, where we replace
\begin{equation}
    S_0(\bf{x}, \bf{y}) \to \frac{\pmb{\sigma}^a (\bf{x}-\bf{y})^a}{2\pi \brac{(\bf{x} - \bf{y}) \cdot (\bf{x} - \bf{y}) + \varepsilon^2}}
\end{equation}
in the integrals. The divergence of the gauge current is shown in figure \ref{fig: min length anomaly} relative to the Laplacian of $\alpha$, with an identical plot found for the chiral current. We see that for this regulator we have
\begin{equation}
    a = b = \frac{1}{2} \ ,
\end{equation}
and so as one may suspect this is not gauge-invariant. Now let us consider Pauli-Villars regularization using the five regulator fields with masses \eqref{eq: c2 and c3} and \eqref{eq: c4 and c5}, where we add a contribution from each to \eqref{eq: anomaly numerics} with alternating signs.
\begin{figure}
    \centering
    \includegraphics[width=0.8\linewidth]{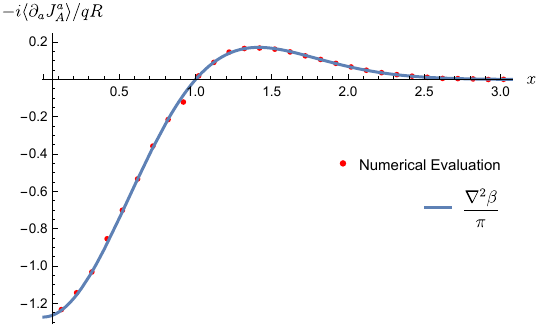}
    \caption{Numerical evaluation of the one-loop contribution to the divergence of $J_A^a$ using Pauli-Villars regularization with $\epsilon= 0.01$; the exact result for the anomaly coefficient $b=1$ is given for comparison.}
    \label{fig: PV chiral anomaly}
\end{figure}
The divergence of the axial current is shown in figure \ref{fig: PV chiral anomaly}, exactly matching the Laplacian of $\beta$ with coefficient $b=1$, meaning $a=0$ (which can also be checked numerically) and we have a gauge-invariant regularization.

\bibliography{references.bib}

\end{document}